\documentclass[a4paper,11pt]{article}
\pdfoutput=1 % if your are submitting a pdflatex (i.e. if you have
\usepackage{jcappub} % for details on the use of the package, please
\usepackage{amsmath}  
\usepackage[T1]{fontenc} % if needed
\usepackage{enumerate}
\usepackage[color=orange]{todonotes}
\usepackage[normalem]{ulem}
\DeclareMathOperator{\csch}{csch}
\title{\boldmath Higgs-Dilaton Inflation in Einstein-Cartan gravity}
\author{Matteo Piani}
\author{and Javier Rubio}
\affiliation{Centro de Astrof\'{\i}sica e Gravita\c c\~ao  - CENTRA,
Departamento de F\'{\i}sica, Instituto Superior T\'ecnico - IST,
Universidade de Lisboa - UL, Av. Rovisco Pais 1, 1049-001 Lisboa, Portugal.}
\emailAdd{matteo.piani@tecnico.ulisboa.pt}
\emailAdd{javier.rubio@tecnico.ulisboa.pt}
\abstract{We study the phenomenology of the Higgs-Dilaton model in the context of Einstein-Cartan gravity, focusing on the separate impact of the Holst and Nieh-Yan terms on the inflationary observables. Using analytical and numerical techniques, we show the predictions of these scenarios to display an attractor-like behaviour intrinsically related to the curvature of the field-space manifold in the metric formulation of the theory. Beyond that, the analysis of the Nieh-Yan case reveals the existence of an additional attractor solution induced by a cubic pole in the inflaton kinetic term that becomes relevant at large dilaton couplings. This constitutes a unique feature of the Einstein-Cartan formulation as compared to the metric and Palatini counterparts.}
\begin{document}
%%%%%%%%%%%%%%%%%%%%%%%%%%%%%%%%%%%%%%%%%%%%%%%%%%%%%%%%%%%%%%%%%%%%%%%%%%%%%%%%
\maketitle
\flushbottom

%%%%%%%%%%%%%%%%%%%%%%%%%%%%%%%%%%%%%%%%%%%%%%%%%%%%%%%%%%%%%%%%%%%%%%%%%%%%%%%%
\section{Introduction} \label{sec:intro}
%%%%%%%%%%%%%%%%%%%%%%%%%%%%%%%%%%%%%%%%%%%%%%%%%%%%%%%%%%%%%%%%%%%%%%%%%%%%%%%%

In the last few decades we have witnessed the transition to an era of precision cosmology, with the most recent data from the Cosmic Microwave Background (CMB)~\cite{Planck:2018jri,BICEP:2021xfz} consolidating inflation~\cite{Guth:1980zm,Linde:1981mu,Mukhanov:1981xt} as the most successful paradigm for the description of the very early Universe. Among the plethora of inflationary models in the literature, Higgs Inflation (HI)~\cite{Bezrukov:2007ep} stands out as a minimal scenario, not requiring new degrees of freedom beyond the Standard Model (SM) content while potentially providing a connection between collider experiments and cosmological observations (for a review, see Ref.~\cite{Rubio:2018ogq}). 
Interestingly enough, the inclusion of a non-minimal coupling of the Higgs field to gravity, together with the unavoidable existence of Standard Model fermions,\footnote{Since fermions source torsion, they can induce changes in some metric-affine theories, namely Einstein-Cartan gravity. However, as long as the action coincides with that in the metric case, these effects are Planck mass suppressed and therefore observationally irrelevant at subplanckian energies \cite{Kibble:1961ba}.} provides also an indirect way of testing the fundamental nature of gravity, explicitly breaking the well-known equivalence of pure metric-affine gravitational theories \cite{Karananas:2021zkl,Diakonov:2011fs,Rigouzzo:2022yan}.

The original HI proposal employed the so-called \textit{metric} or \textit{Hilbert formulation}~\cite{Bezrukov:2007ep}. Subsequently, the scenario has been studied in \textit{Palatini}~\cite{Bauer:2008zj},  teleparallel \cite{Raatikainen:2019qey} and affine \cite{Azri:2017uor} formulations of gravity. In the metric and Palatini cases, the curvature tensor $R_{\sigma\mu\nu}^{\rho}$ is expressed in terms of a symmetric connection $\Gamma^{\rho}_{\nu\sigma}$,
with the only difference laying on the choice of fundamental degrees of freedom. In particular, while in the metric formulation the connection is taken to be the usual Levi-Civita one, and thus defined through the metric $g_{\mu\nu}$, these two quantities are assumed to be independent in the Palatini formulation of gravity, being their relation only recovered through the equations of motion.
When it comes to HI both scenarios share the same set up; the Standard Model action is supplemented with a non-minimal coupling of the Higgs field $H$ to the scalar curvature $R=g^{\mu\nu}R_{\mu\nu}$, namely 
\begin{equation}\label{Eq:EH-non-minimal}
    S_{\text{grav}} =\frac{1}{2}\int d^4x\sqrt{-g}\:(M_P^2+\xi H^\dagger H)R\,,
\end{equation}
with  $M_P=2.44\times 10^{18}\,{\rm GeV}$  the reduced Planck mass and $\xi$ a dimensionless coupling constant taken to be positive definite and significantly larger than one.
From the phenomenological point of view, the two scenarios differ on the efficiency of the heating stage \cite{Garcia-Bellido:2008ycs,Bezrukov:2008ut,Bezrukov:2014ipa,Rubio:2015zia,Repond:2016sol,Ema:2016dny,DeCross:2016cbs,Sfakianakis:2018lzf,Rubio:2019ypq} and the value of the tensor-to-scalar ratio $r$ characterizing the relative amplitude of tensor and curvature perturbations generated during inflation \cite{Bezrukov:2007ep,Bezrukov:2017dyv,Bauer:2008zj,Rasanen:2017ivk,Shaposhnikov:2020fdv}. While  in the metric formulation the latter quantity is independent of the non-minimal coupling to gravity and within the reach of upcoming CMB polarization experiments \cite{CMB-S4:2016ple,Hazumi:2019lys}, $r\sim 10^{-3}$, it becomes $\xi$-dependent and generally much smaller than one in the Palatini case,  $r\sim 10^{-13}\ldots 10^{-4}$, not leading therefore to a detectable gravitational waves signal.
From the formal point of view, the two formulations display also a different range of validity. In particular, while the inflationary plateau is always of the order ${\cal O} (M_P/\xi)$, the associated cut-off scales signaling the potential violation of tree-level unitarity are significantly different for fluctuations computed around the electroweak vacuum \cite{Barbon:2009ya,Burgess:2009ea,Burgess:2010zq,Hertzberg:2010dc,Bauer:2010jg,Ito:2021ssc,Antoniadis:2022ogv,Karananas:2022byw}, $\Lambda_\text{metric}\sim M_P/\xi$ and $\Lambda_\text{Palatini}\sim M_P/\sqrt{\xi}$,  being the Palatini one parametrically larger. This reduces the uncertainties associated to the non-renormalizability of the Standard Model non-minimally coupled to gravity and eases the connection between inflationary observables and collider experiments in Palatini Higgs inflation \cite{Shaposhnikov:2020fdv}.

Recent works~\cite{Langvik:2020nrs,Shaposhnikov:2020gts,Karananas:2021zkl} have extended the metric and Palatini versions of HI to Einstein-Cartan (EC) gravity~\cite{Kibble:1961ba,Utiyama:1956sy}. As compared to the Palatini case, the connection in EC gravity, while still independent of the metric, is no longer symmetric with respect to the exchange of lower indices, giving rise to a non-vanishing torsion contribution $T^{\rho}_{\nu{}\sigma}=\Gamma^{\rho}_{\nu{}\sigma}-\Gamma^{\rho}_{\sigma{}\nu}$, with $T^{\rho}_{\nu{}\sigma}$ the torsion tensor. This allows for the inclusion of extra terms in the action and consequently for additional non-minimal couplings of the Higgs field to gravity, making it possible to obtain sizable tensor-to-scalar ratios slightly below the current observational bound \cite{Shaposhnikov:2020gts}. This constitutes a novel feature of EC HI as compared with the metric and Palatini formulations, where large values of this quantity appear only after the inclusion of quantum effects and for rather fine-tuned corners of parameter space \cite{Bezrukov:2014bra,Hamada:2014iga,Bezrukov:2017dyv}. On top of that, since the torsion is sourced by the fermions in the SM and beyond, these species can contribute to the dynamics of the theory through non-minimal kinetic terms \cite{Freidel:2005sn,Alexandrov:2008iy,Diakonov:2011fs}, effectively inducing higher-dimensional Higgs-fermion and fermion-fermion interactions with a potential impact on the heating stage and the production of fermionic dark matter \cite{Shaposhnikov:2020aen}. All in all, EC Higgs Inflation displays a far richer inflationary phenomenology than the metric and Palatini counterparts, while incorporating these scenarios as limiting cases ~\cite{Shaposhnikov:2020gts,Langvik:2020nrs}. Besides that,  EC gravity can be seen as the gauge theory of the Poincar\'e group~\cite{Kibble:1961ba}, establishing an appealing parallelism between gravity and the other elementary forces in Nature. 

Despite the very attractive prospective of having the only known fundamental scalar playing the role of the inflaton, one must keep in mind that the Higgs boson is not free from issues. On the one hand, it is reasonable to wonder about the origin of the wide separation between the Fermi and the Planck scales. On the other hand, the Higgs mass is supposed to be highly unstable under radiative corrections~\cite{Giudice:2008bi}, requiring therefore a large fine-tuning of the tree-level mass parameter in order to reproduce the value observed at collider experiments (see, however, Refs.~\cite{Mooij:2021ojy,Manohar:2018aog} for counter arguments). A possible line of attack to this  well-known \textit{hierarchy problem} is to consider a scale-invariant extension of the SM non-minimally coupled to gravity~\cite{Garcia-Bellido:2011kqb,Rubio:2020zht}, i.e. invariant under global dilatations. This symmetry forbids not only the presence of a Higgs mass parameter in the Standard Model action but also the inclusion of the infamous cosmological constant and the Planck mass term in Eq.~\eqref{Eq:EH-non-minimal}, thus removing every explicit mass scale from the classical formulation of the theory.

In the so-called \textit{Higgs-Dilaton model} \cite{Shaposhnikov:2008xb,Garcia-Bellido:2011kqb,Bezrukov:2012hx} (see Ref.~\cite{Rubio:2020zht} for a review), all physical scales are generated dynamically through the spontaneous symmetry breaking of dilatations. This process is triggered by a dilaton field, an extra singlet under the Standard Model gauge group that provides the order parameter for dilatations. Assuming  scale invariance to remain exact at quantum level and no new degrees of freedom to appear below the Planck scale, the Higgs mass will not be sensitive to quantum corrections  \cite{Shaposhnikov:2008xi,Garcia-Bellido:2011kqb,Bezrukov:2012hx,Armillis:2013wya,Gretsch:2013ooa}. This relaxes the stability part of the hierarchy problem, not providing, however, a direct explanation for the many orders of magnitude separating the different scales. At the same time, the  absence of perturbative quantum corrections opens the possibility of generating mass-splittings through non-perturbative gravitational effects. This appealing scenario was put forward in Ref.~\cite{Shaposhnikov:2018xkv} and subsequently extended to different scale-invariant settings \cite{Shaposhnikov:2018jag,Shkerin:2019mmu,Karananas:2020qkp}. 

Inflation in the Higgs-Dilaton model has been extensively studied in the metric and Palatini formulations (see Refs.~\cite{Garcia-Bellido:2011kqb,Bezrukov:2012hx,Rubio:2014wta,Garcia-Bellido:2012npk,Casas:2017wjh,Herrero-Valea:2019hde,Karananas:2020qkp} and Ref.~\cite{Rubio:2020zht} for a review).
In the context of EC gravity the general framework was introduced in Refs.~\cite{Shaposhnikov:2020frq,Karananas:2021gco}, but its phenomenological consequences are still to be analyzed. In the presence of torsion, the associated Lagrangian contains many gravitational operators that can be non-minimally coupled to the scalar fields, even if the interaction terms are restricted to operators of mass dimension not greater than four. The goal of this work is to study the phenomenological implications and viable parameter space of a restricted EC scenario including only the well-known Holst and Nieh-Yan operators, as first assumed in Refs.~\cite{Langvik:2020nrs,Shaposhnikov:2020gts,Shaposhnikov:2020frq}. These seminal operators play an important role in Loop Quantum Gravity approaches and the study of black hole entropy (see e.g. Ref.~\cite{Langvik:2020nrs} and references therein). On top of that, they come together with some interesting phenomenology. On the one hand, the Holst term can be responsible for the appearance of parity violating interactions between fermions~\cite{Shapiro:2014kma}, which in turn could lead to observable quantum effects. On the other hand, the presence of the Nieh-Yan term turns out to reproduce both the metric and the Palatini versions of the theory, connecting them in a smooth way. 

Our analysis is intended to be strictly classical, leaving the discussion of quantum effects to a future work. The inclusion of these radiative corrections is expected to restrict the viable parameter space of the model by selecting a preferred window for the Higgs self-interaction through the renormalization group running, limiting with it the range of non-minimal couplings to gravity able to reproduce the observed amplitude of CMB anisotropies. Further restrictions could follow from demanding the unknown UV completion to incorporate important principles for our understanding of Nature, such as unitarity, causality, locality and Lorentz invariance \cite{Herrero-Valea:2019hde}. The most typical example is the 2-to-2 scattering amplitude in the forward limit, which, combined with the above requirements, forces the Wilson coefficients to exceed a given value that can be computed without any knowledge of the UV completion \cite{Adams:2006sv,Bellazzini:2017fep}.

This paper is organized as follows. In Section~\ref{sec:HDSI} we introduce the prototypical Higgs-Dilaton inflation action, complementing it with the aforementioned Holst and Nieh-Yan terms. Switching the resulting model to the Einstein frame and using the equations of motion, we obtain then an equivalent metric formulation of the theory that we use to highlight some important properties permeating the subsequent analysis. The individual impact of the Holst and Nieh-Yan terms on the inflationary observables is considered in Section~\ref{sec:Numerical}, where we make use of an ensemble Markov chain Monte Carlo (MCMC) method to sample the corresponding parameter space. In both cases, we benefit from an analytical approximation in which the Gauss curvature of the field manifold becomes approximately constant during inflation, arguing that this maximally symmetric regime leads to an exponential stretching of the canonically normalized inflaton field able to desensitize the spectral properties from the shape of the potential. Finally, in Section \ref{sec:conclusions}, we present our conclusions, outlining the directions for future research.

%%%%%%%%%%%%%%%%%%%%%%%%%%%%%%%%%%%%%%%%%%%%%%%%%%%%%%%%%%%%%%%%%%%%%%%%%%%%%%%%
\section{Higgs-Dilaton model with the Holst and Nieh-Yan term}\label{sec:HDSI}
%%%%%%%%%%%%%%%%%%%%%%%%%%%%%%%%%%%%%%%%%%%%%%%%%%%%%%%%%%%%%%%%%%%%%%%%%%%%%%%%

The fundamental degrees of freedom in EC gravity are the tetrad and the spin connection, being the metric constructed out of the former. In this paper, we will be interested in scenarios not leading to additional degrees of freedom beyond the SM content, the dilaton and the massless graviton. To this end, we follow the guidelines of Ref.~\cite{Karananas:2021gco}, restricting ourselves to a renormalizable matter action in the flat spacetime limit and accounting only for gravitational operators of mass dimension not greater than two. Among them, we focus our attention on two special terms:  the Holst term~\cite{PhysRevD.22.1915,Nelson:1980ph,Castellani:1991et} and the Nieh-Yan topological invariant~\cite{Nieh:1981ww,Nieh:2008btw}. Further restricting the coupling between matter and gravity to operators of mass dimension smaller than four, the gravitational part of the theory becomes the sum of three separate pieces,
\begin{equation}\label{eq:fulltheory}
S=S_{\text{0}}+S_{\text{Holst}}+S_{\text{NY}}\,,   
\end{equation}
 whose precise formulation in terms of the tetrad and the spin connection can be found Appendix \ref{appendix:Derivation}. The first term stands for the prototype Higgs-Dilaton action,
\begin{equation}\label{EH-action}
    S_0 =\int d^4x\sqrt{-g}\,\left[\frac{\tau \chi^2+\xi h^2}{2}R-\frac{1}{2}\left(\partial h\right)^{2}-\frac{1}{2}\left(\partial \chi\right)^{2}-U(\chi, h)\right]\,,
\end{equation}
with $h$ the Higgs field in the unitary gauge $H=(0,h/\sqrt{2})^T$, $\chi$ the dilaton field, $\tau$ a dimensionless coupling constant and
\begin{equation}\label{SSB-potential}
    U(\chi,h)=\frac{\lambda}{4}\left( h^2-\alpha \chi^2 \right)^2+\beta \chi^4
\end{equation}
a scale-invariant potential, with $\lambda$ the Higgs self-coupling and $\alpha$ and $\beta$ two small parameters set to reproduce, respectively, the value of the Fermi scale and the cosmological constant in Planckian units. For vanishing $\beta=0$, the theory \eqref{EH-action} displays a flat-space limit and a continuous family of symmetry-breaking ground states, $\langle h \rangle^2=\alpha \langle \chi \rangle^2$.  For $\beta\neq 0$, the spontaneous symmetry breaking generates also a cosmological constant term, while only slightly modifying the previous vacuum manifold if $\beta \ll \alpha\ll 1$ and $\beta, \alpha \ll \tau,\xi$ \cite{Garcia-Bellido:2011kqb}, a parameter hierarchy that we will assume in what follows.~\footnote{A dynamical dark energy component can also appear if the invariance under dilatations is explicitly broken in a controllable way \cite{Wetterich:1987fm,Rubio:2017gty}. A simple possibility, pointed out in Refs.~\cite{Shaposhnikov:2008xb,Blas:2011ac,Garcia-Bellido:2011kqb}, is to restrict the spacetime diffeomorphisms invariance to the minimal group of transverse diffeomorphisms leading to spin-two gravitons \cite{vanderBij:1981ym,Buchmuller:1988wx,Alvarez:2006uu}. Interestingly enough, this setting involves potentially-testable consistency relations between the inflationary and dark energy observables \cite{Garcia-Bellido:2011kqb,Trashorras:2016azl,Casas:2017wjh,Casas:2018fum}. }

The second  term, 
\begin{equation}\label{Holst-action}
   S_{\text{Holst}}= \frac{1}{2\bar{\gamma}}\int d^4x\sqrt{-g}\:(\tau \chi^2+\xi_\gamma h^2)\varepsilon^{\mu\nu\rho\sigma}R_{\mu\nu\rho\sigma}\,,
\end{equation}
is associated with the loss of symmetries of the curvature tensor $R_{\mu\nu\rho\sigma}$ in EC gravity, which allows for its contraction with the antisymmetric tensor $\varepsilon^{\mu\nu\rho\sigma}$, with $\varepsilon_{0123}=1$. Note that, in spite of the appearances, this expression contains just two physical parameters, since the global Barbero-Immirzi parameter $\bar{\gamma}$ ~\cite{Immirzi:1996dr,Immirzi:1996di} could be always reabsorbed into the definition of the individual non-minimal couplings $\tau$ and $\xi_\gamma$. The chosen normalization aims at facilitating the comparison with Ref.~\cite{Shaposhnikov:2020frq}, where the dilaton coupling is taken to be the same as in Eq.~(\ref{EH-action}) and the second non-minimal coupling $\xi_\gamma$ is assumed to be \textit{a priori} independent from $\xi$. 

Finally, the third Nieh-Yan term is given by
\begin{equation}\label{Nieh-Yan-Action}
   S_{\text{NY}} =\frac{1}{2}\int d^4x\:(\tau_\eta \chi^2+\xi_\eta h^2)\partial_{\mu}\left(\sqrt{-g}\epsilon^{\mu\nu\rho\sigma}T_{\nu\rho\sigma}\right)\,,
\end{equation}
where we have introduced two additional non-minimal couplings $\tau_\eta$ and $\xi_\eta$ for the dilaton and the Higgs field. Note that, due to these couplings, this piece is no longer a boundary term, but rather contributes to the equations of motion. 

The full EC theory under consideration can be recast as a purely metric theory, more suitable for phenomenological analyses. To obtain this, the first step is switching the action \eqref{eq:fulltheory} to the Einstein frame in order to get rid of the non-minimal coupling to the scalar curvature $R$. To this end, we choose to work in Planckian units $M_P=1$ and perform a Weyl rescaling of the metric, $g_{\mu \nu} \rightarrow \Omega^{-2} g_{\mu \nu}$, with conformal factor ~\cite{Garcia-Bellido:2011kqb} 
\begin{equation}\label{Conf-h}
\Omega^{2}=\tau \chi^2+\xi h^{2}\,.
\end{equation}
Then, we remove the dependence on torsion by splitting the connection into its torsionless and torsionful parts, solving subsequently for the latter and plugging back the solution into the action (see Refs.~\cite{Shaposhnikov:2020frq,Karananas:2021zkl} for an explicit computation). In this way, we end up with an equivalent metric theory,
 \begin{equation}\label{Eq:Final-Action}
    S= \int d^4x\sqrt{-g} \left[\frac{R}{2} - \frac{1}{2}g^{\mu\nu}K_{ab}\partial_\mu \varphi^a \partial_\nu \varphi^b - V(\varphi)\right],\hspace{10mm}V(\varphi)=\frac{U(\varphi)}{\Omega^4}\,,
\end{equation}
 where we have collectively denoted the scalar fields as $\varphi^a=(\chi,h)^T$ and defined a field-space metric
\begin{equation}\label{K-matrix}
    K_{ab}=  \dfrac{1}{\Omega^2}\left[ \begin{array}{cc}
  1+  6  \tau^2 \dfrac{(\gamma - \bar{\tau})^2}{\gamma^2+1} \dfrac{\chi^2}{\Omega^2}\,\,\,    & 6\tau \xi\dfrac{(\gamma - \bar{\xi})(\gamma - \bar{\tau})}{\gamma^2+1}  \dfrac{\chi h}{\Omega^2}  \\
  6\tau \xi \dfrac{(\gamma - \bar{\xi})(\gamma - \bar{\tau})}{\gamma^2+1}  \frac{\chi h}{\Omega^2} \,\,\,  & 1+   6  \xi^2 \dfrac{(\gamma - \bar{\xi})^2}{\gamma^2+1} \frac{h^2}{\Omega^2} 
     \end{array}\right]\,,
\end{equation}
with
\begin{equation}
        \gamma(\chi,h)=\frac{\tau \chi^2 +\xi_\gamma h^2}{\bar{\gamma}\Omega^2}, \hspace{15mm}\bar{\tau}= \frac{\tau_\eta + \frac{\tau}{\bar{\gamma}}}{\tau}\,,
        \hspace{15mm}\bar{\xi}= \frac{\xi_\eta + \frac{\xi_\gamma}{\bar{\gamma}}}{\xi}\,.
\end{equation}
We are interested in scenarios where this model can support inflation, and in particular where the inflationary phase is mainly driven by the SM Higgs. Several comments are in order at this point:  
\begin{enumerate}
\item In the metric form \eqref{Eq:Final-Action}, the non-minimal couplings to gravity are traded for a specific set of higher-dimensional operators. The consistency of the theory requires the eigenvalues of the field-space metric \eqref{K-matrix} to be positive definite \cite{Shaposhnikov:2020frq}. This can be trivially achieved for non-negative couplings, as we assume in what follows. More general choices are \textit{a priori} possible \cite{Langvik:2020nrs}, and remain to be tested in this scenario. Note, however, that albeit potentially giving rise to valid inflationary solutions\footnote{For $\xi<0$, the conformal factor \eqref{Conf-h} goes to zero at $h=\sqrt{\tau/\xi} \,\,\chi$, leading to an attractor behaviour.}~\cite{Langvik:2020nrs}, some choices like $\xi<0$ can be automatically excluded by requiring the aforementioned eigenvalues to remain positive for all possible field configurations. 

\item The two-dimensional field-space metric \eqref{K-matrix} is fully characterized by its Gaussian curvature. In the large $h$ limit, and at leading order in the dilaton couplings, this quantity becomes approximately constant, 
\begin{equation}\label{eq:gencurvature}
\kappa=
\frac{(\tau -\xi )
   \left(\xi^2+\frac{\xi_\gamma ^2}{{\bar\gamma}^2}\right)^2+ 12\xi ^2 \left(\frac{\xi_\eta ^2
   \tau }{\xi_\gamma }-\xi_\eta  \tau_\eta
   \right)\frac{ \xi_\gamma ^2}{{\bar\gamma}^2}-12 \xi ^3 \xi_\eta  (\xi  \tau_\eta -\xi_\eta  \tau )}{\left(\xi ^2+\frac{\xi_\gamma ^2}{\bar\gamma^2}\right) \left(\xi^2+
  \frac{\xi_\gamma
   ^2}{{\bar\gamma}^2} +6\xi  \xi_\eta^2\right)}\,,
\end{equation}
making the associated manifold maximally symmetric. As argued in Refs.~\cite{Karananas:2016kyt,Casas:2018fum,Rubio:2020zht,Karananas:2020qkp}, this property is expected to determine, almost completely, the inflationary observables \textit{in the associated regime}, provided, of course, that the amplitude of the potential satisfies the CMB normalization \cite{Planck:2018jri}. 

Note that the metric and Palatini formulations summarized in Ref.~\cite{Rubio:2020zht} are well-captured by the EC kinetic structure. In particular, in the Holst-like inflation limit, $\xi_\eta=\tau_\eta= 0$, the curvature \eqref{eq:gencurvature} coincides with the Palatini one, 
\begin{equation}\label{eq:kH}
    \kappa_H=-\xi\left(1-\frac{\tau}{\xi}\right)\,,
\end{equation}
while in the complementary Nieh-Yan-like inflation one, $\bar{\gamma}\rightarrow \infty,\;\xi_\gamma=0$, it becomes
\begin{equation}\label{NY-curvature}
\kappa_{NY}= -\frac{\xi 
   (\xi -\tau )+12 \xi_\eta  (\xi  \tau_\eta -\xi_\eta  \tau )}{\xi +6 \xi_\eta ^2}\,,
\end{equation}
reducing therefore to the metric value $\kappa\simeq -1/6$ for $\tau_\eta=\tau,\;\xi_\eta=\xi$. Needless to say, the latter case is also recovered for vanishing Holst and Nieh-Yan terms, $\bar \gamma=\xi_\gamma=\xi_\eta =\tau_\eta=0$. Therefore, by continuously varying the curvature of the field manifold, we can transition from the metric to the Palatini formulation, while potentially exploring new phenomenological regimes. 

\item Although it might seem that the action (\ref{Eq:Final-Action}) gives rise to a multi-field inflationary scenario, we can exploit scale invariance to reduce the dynamics to that of a single field model. This particular feature can be extended to other multi-field models displaying alternative continuous symmetries like $U(1)$ or shift-invariance, as done for instance in Refs.~\cite{Greenwood:2012aj,Kaiser:2013sna,Achucarro:2019pux}. For the HD model this reduction can be easily done by inspecting the Noether's current associated to dilatations~\cite{Garcia-Bellido:2011kqb,Casas:2018fum,Ferreira:2016wem}, as worked out explicitly in Appendix \ref{appendix:diagonalization}.  In particular, it is always possible to choose a set of variables $(\rho,\theta)$ reducing the action \eqref{Eq:Final-Action} to the diagonalized form \cite{Blas:2011ac,Karananas:2016kyt,Rubio:2020zht}
\begin{equation}\label{SI-Lagrangian}
    S=\int d^4x\sqrt{-g} \left[\frac{R}{2}-\frac{1}{2} K_\theta(\theta)(\partial\theta)^2-\frac{1}{2}K_\rho(\theta)(\partial \rho)^2-V(\theta) \right]\,,
\end{equation}
where $\theta$ is a function of $h/\chi$ only (and therefore scale invariant) and $\rho$ a physical dilaton appearing only through derivatives. The action \eqref{SI-Lagrangian} comes along with some phenomenological consequences. On the one hand, having reduced the dynamics to that of a single field scenario leads to the absence of large isocurvature perturbations and sizable non-gaussianities~\cite{Garcia-Bellido:2011kqb,Ferreira:2018qss,Rubio:2020zht}. On the other hand, being the dilaton a Goldstone boson, it displays only derivative couplings to matter, thus preventing the presence of unobserved fifth forces among the SM constituents \cite{Garcia-Bellido:2011kqb,Ferreira:2016kxi,Burrage:2018dvt,Rubio:2020zht}.
Moreover, one can further exploit the aforementioned conservation law to show that the condition $d \rho/dN=0$, with $N$ the number of $e$-folds of inflation, represents an attractor solution constraining the motion on the $(\chi,h)$ plane~\cite{Garcia-Bellido:2011kqb,Ferreira:2016wem,Casas:2018fum,Rubio:2020zht}. 
\end{enumerate}

%%%%%%%%%%%%%%%%%%%%%%%%%%%%%%%%%%%%%%%%%%%%%%%%%%%%%%%%%%%%%%%%%%%%%%%%%%%%%%%%
\section{Numerical Analysis}\label{sec:Numerical}
%%%%%%%%%%%%%%%%%%%%%%%%%%%%%%%%%%%%%%%%%%%%%%%%%%%%%%%%%%%%%%%%%%%%%%%%%%%%%%%%

The standard procedure for the analysis of the model \eqref{SI-Lagrangian} would be to canonically normalize the inflaton field to subsequently compute the inflationary observables in terms of a normalized variable $\tilde \theta$,
\begin{equation}\label{Eq:normal-field}
    \frac{d\tilde{\theta}}{d \theta}= \sqrt{K_{\theta}} \;\hspace{5mm} \Longrightarrow \hspace{5mm}    \tilde{\theta}=\int \sqrt{K_\theta(\theta)} \,d \theta\,.
\end{equation}
However, the functional form of $K_\theta$ makes impractical to solve \eqref{Eq:normal-field} or to invert its solution in order to obtain an analytical expression for $\theta(\tilde{\theta})$. To overcome this difficulty, we will perform a numerical MCMC scanning of the parameter space in non-canonical variables, computing the tensor-to-scalar ratio, the spectral tilt, its running and the running of the running in the slow-roll approximation. These observables are respectively given by~\cite{Planck:2018jri}
\begin{eqnarray}\label{ns-r-SR}
    r &=& 16\epsilon_V\,,\\
    n_s &=& 1-6\epsilon_V+2\eta_V\,, \\ 
    \alpha_s &=& -24\epsilon_V^2+16\epsilon_V\eta_V-2\xi_V^2\,,\\
    \beta_s &=& -192\epsilon_V^3+192\epsilon_V^2\eta_V-32\epsilon_V\eta_V^2-24\epsilon_V \xi_V^2+2\eta_V\xi_V^2+2\varpi_V^3\,, \label{ns-r-SR2}
\end{eqnarray}
with
\begin{equation}\label{inflationary-parameters-general}
\epsilon_V=\frac{1}{2}\left( \frac{V_{,\tilde{\theta}}}{V(\tilde{\theta})}\right)^2\,, \hspace{10mm}
\eta_V= \frac{V_{,\tilde{\theta}\tilde{\theta}}}{V(\tilde{\theta})}\,,\hspace{10mm} \xi_V^2=\frac{V_{,\tilde{\theta}}}{V^2(\tilde{\theta})}V_{,\tilde{\theta}\tilde{\theta}\tilde{\theta}}\,,\hspace{10mm}
\varpi_V^3=2\epsilon_V  \frac{V_{,\tilde{\theta}\tilde{\theta}\tilde{\theta}\tilde{\theta}}}{V(\tilde{\theta})}\,,\end{equation}
the usual slow-roll parameters and the derivatives with respect to the canonical field $\tilde{\theta}$ implicitly computed in terms of the non-canonical variable $\theta$, cf.~Eq.~\eqref{Eq:normal-field}.

In a similar fashion to Ref.~\cite{Shaposhnikov:2020gts}, we look for positive values of the couplings leading to a given number of $e$-folds 
\begin{equation}\label{NEFOLDS}
N=\int^{\tilde{\theta}^*}_{\tilde{\theta}_E}\frac{d \tilde{\theta}}{\sqrt{2\epsilon_V(\tilde{\theta})}}= \int^{\theta^*}_{\theta_{E}} \sqrt{\frac{{K_\theta}}{{2\epsilon_V(\theta)}}}  d \theta\,,
\end{equation}
with the integration in this expression performed from the value of the field at the end of inflation, $\theta_{E}$, to the one at the horizon exit of a given pivot scale, $\theta^*$.  The value at horizon exit is fixed by the CMB normalization $V(\theta_*)/\epsilon_V(\theta_*)= 5.0 \cdot 10^{-7}$  \cite{Planck:2018jri}, while that at the end of inflation follows directly from the violation of the slow-roll conditions, namely $\theta_{\text{E}}=\text{max}\left[\theta_\epsilon,\theta_\eta\right]$, with $\theta_\epsilon$ and $\theta_\eta$ satisfying respectively $\epsilon_V(\theta_\epsilon)\simeq 1$ and $|\eta_V(\theta_\eta)| \simeq 1$. The exact value of $N$ depends on the post-inflationary evolution of the Universe, and in particular on the details of the heating stage. Since the study of this epoch in EC gravity goes beyond the purpose of this paper, we choose to fix $N=55$ in what follows, referring the interested reader to Refs.~\cite{Garcia-Bellido:2008ycs,Bezrukov:2008ut,Bezrukov:2014ipa,Rubio:2015zia,Repond:2016sol,Ema:2016dny,DeCross:2016cbs,Sfakianakis:2018lzf,Rubio:2019ypq,Dux:2022kuk} for detailed studies of the limiting metric and Palatini cases.
A similar argument holds for the Higgs self-coupling at the inflationary scale, that we set fiducially to $\lambda=10^{-3}$. Indeed, depending on the low-energy value of the top quark Yukawa coupling, this quantity is expected to vary between $-0.01$ and $0.01$  \cite{Bezrukov:2014ina}. The value $\lambda=10^{-3}$ is therefore well within the latter interval, while being also the preferred one in Palatini Higgs inflation  \cite{Shaposhnikov:2020fdv}.  Intriguingly enough, the recent LHC measurements of the \textit{top pole mass} point also towards this choice \cite{ATLAS:2019guf,CMS:2019esx}.

In order to facilitate the comparison with the previous studies in the literature, we consider in what follows the individual impact of the Holst and Nieh-Yan terms on the prototypical Higgs-Dilaton action \eqref{EH-action}. Since the resulting scenarios involve four free parameters, the constraint on the duration of inflation will determine a three-dimensional hypersurface.  For the sampling of the parameter space we use the sampler provided by the MCMC code \textbf{emcee}~\cite{Foreman-Mackey:2012any}, with a log-likelihood function
\begin{equation}\label{likelihood}
    \ln(L)= - \frac{(55-N_{\text{sample}})^2}{\sigma^2}\,,
\end{equation}
and $N_{\text{sample}}$ the number of $e$-folds obtained from Eq.~(\ref{NEFOLDS}) using the set of parameters given by the sampler. The value of $\sigma$ has no physical meaning and only determines how close $N_{\text{sample}}$ is to 55. Since we are mainly interested in determining regions in parameter space associated with attractor behaviours in the $n_s-r$  plane, we will restrict ourselves to a small value $\sigma=0.01$ leading to hypersurfaces with a fixed number of $e$-folds.  The implications of relaxing this assumption are discussed in Section \ref{sec:conclusions}.

Once the parameter space is scanned, we can compute the inflationary observables in Eqs.~\eqref{ns-r-SR}-\eqref{ns-r-SR2}, with all the quantities there evaluated at horizon exit. Since some of the explored values are related to non slow-roll regimes, we will further restrict $0.7<n_s<1.2$ in order to avoid the sampling of points far away from attractor behaviours.

%%%%%%%%%%%%%%%%%%%%%%%%%%%%%%%%%%%%%%%%%%%%%%%%%%%%%%%%%%%%%%%%%%%%%%%%%%%%%%%%
 \subsection{Holst inflation}\label{sec:Holst}
%%%%%%%%%%%%%%%%%%%%%%%%%%%%%%%%%%%%%%%%%%%%%%%%%%%%%%%%%%%%%%%%%%%%%%%%%%%%%%%%

\begin{figure}
    \centering
    \includegraphics[scale=0.9]{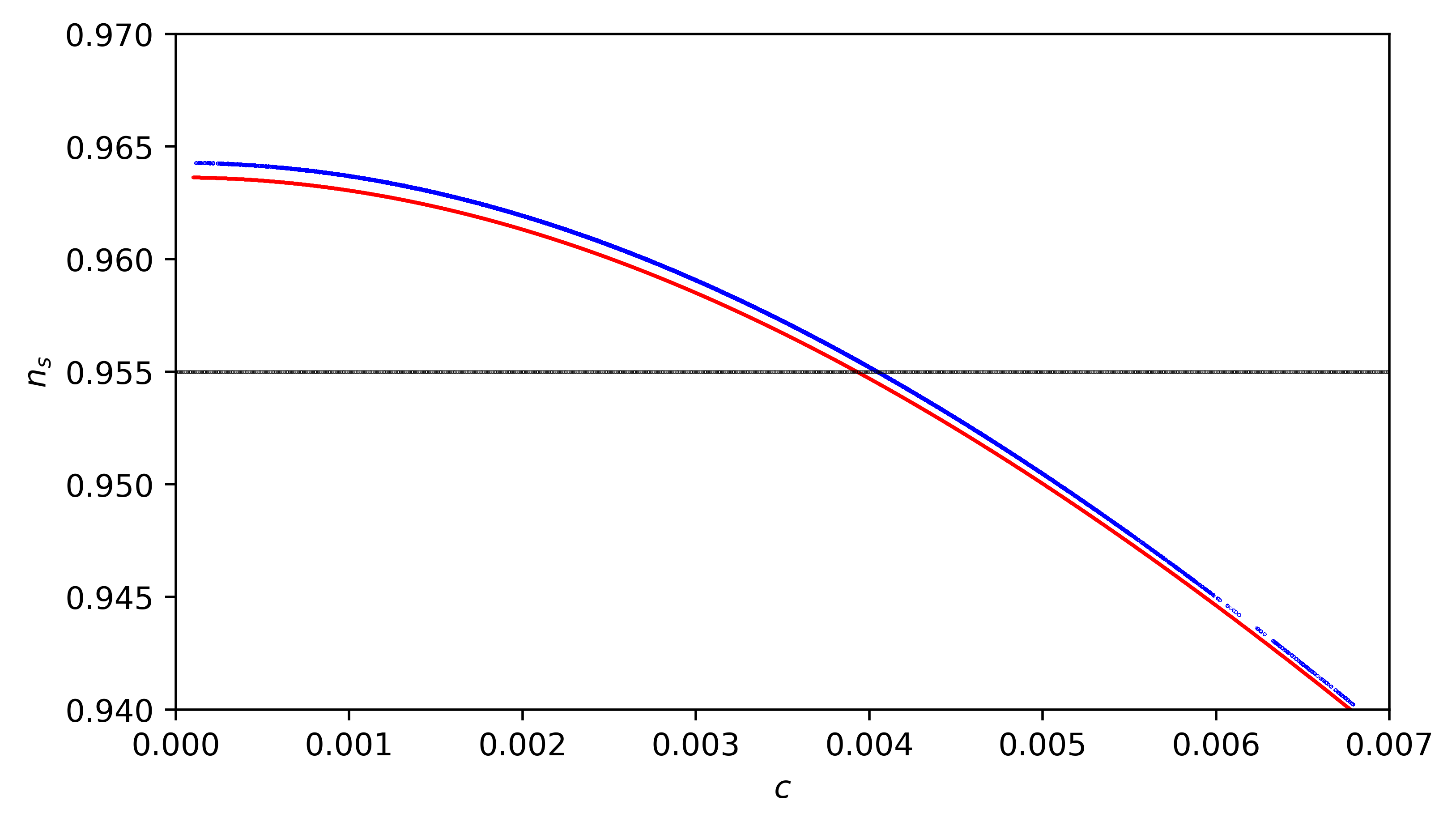}
    \caption{Comparison between the numerical values of $n_s$  (blue line) and the analytical approximation \eqref{ns-r-polestotal} (red line). The sampling is restricted to the maximally symmetric region of the parameter space, namely that at $\xi\sim {\cal O}(10^7)$, with all other parameters left unrestricted. The black horizontal line indicates the lower observational bound on $n_s$ at the $95\%$ C.L.~\cite{BICEP:2021xfz}. }\label{Fig:NS-C}
\end{figure}

\begin{figure}
    \begin{center}
        \fbox{\begin{minipage}[ht]{0.42\linewidth}
            \center{\includegraphics[width=\textwidth]{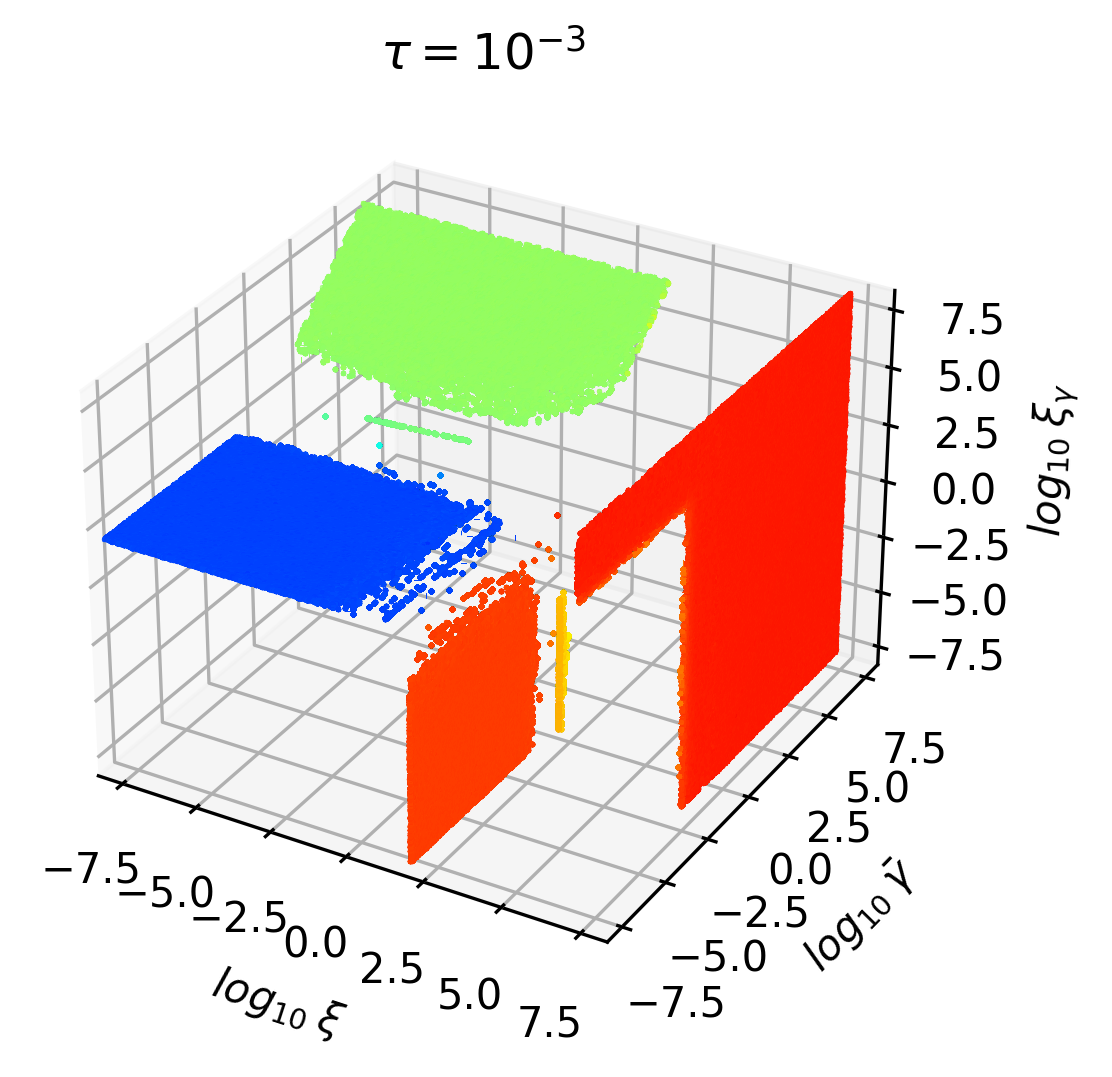} \\(a)}
        \end{minipage}}
               \fbox{\begin{minipage}[ht]{0.42\linewidth}
            \center{\includegraphics[width=\textwidth]{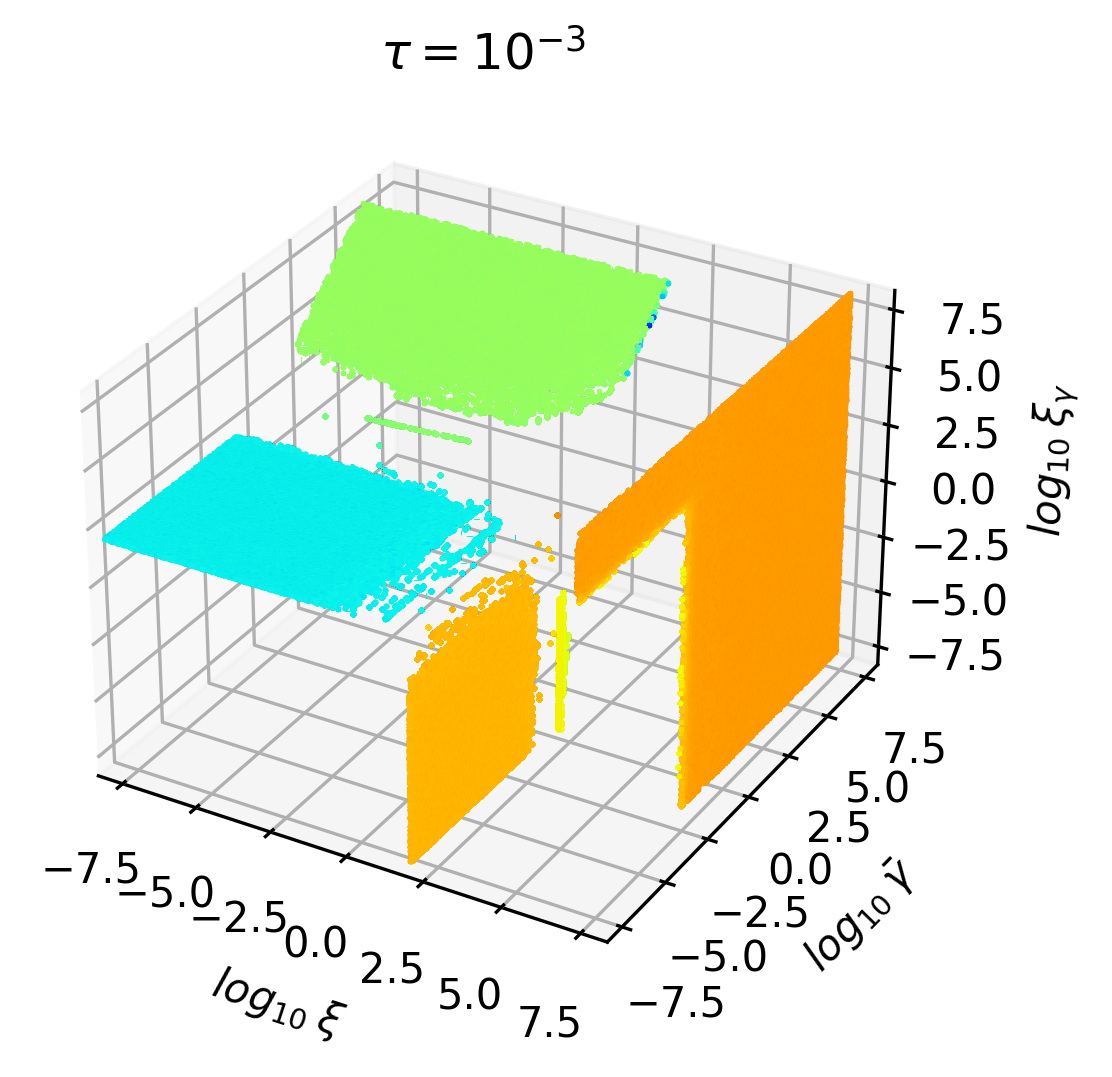} \\(b)}
        \end{minipage}}
                          \begin{minipage}[ht]{0.42\linewidth}
            \center{\includegraphics[width=\textwidth]{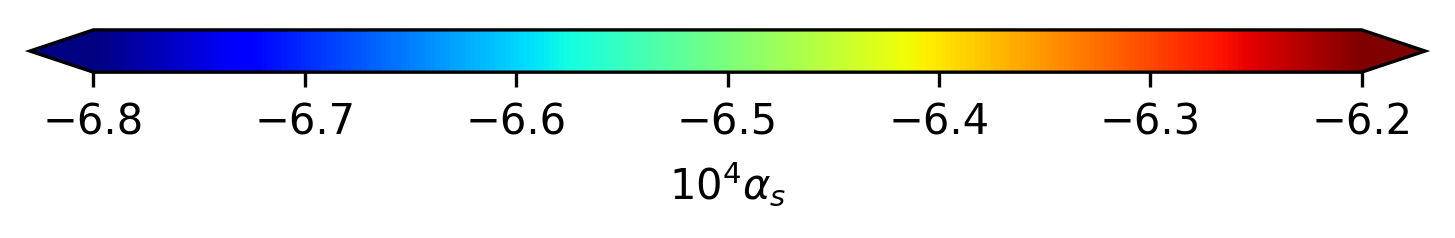}}
             \end{minipage}
                \begin{minipage}[ht]{0.42\linewidth}
            \center{\includegraphics[width=\textwidth]{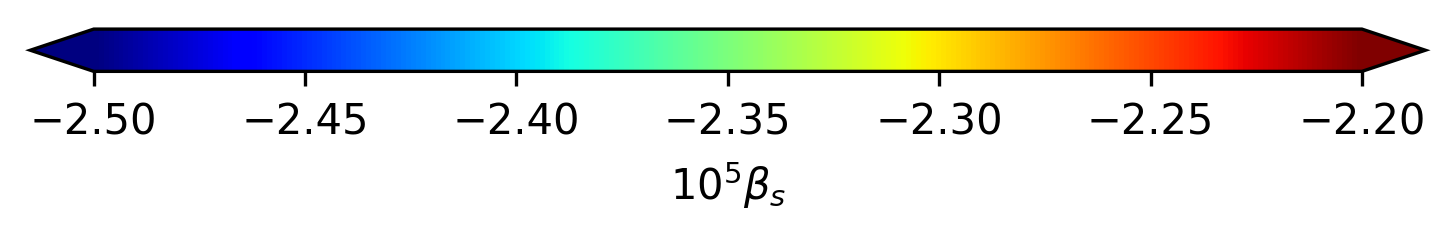}}
             \end{minipage}
    \caption{Numerical results for the running of the scalar spectral tilt  (left panels) and the running of the running (right panels) for a fixed value of the dilaton coupling $\tau=10^{-3}$.} \label{fig:runningsH}
    \end{center} 
\end{figure}

We first turn our attention to the study of Holst-like inflation, obtained by setting $\xi_\eta=\tau_\eta=0$ in Eq.~\eqref{K-matrix},
 \begin{equation}\label{NY-Holst}
  K_{ab}= \frac{1}{\Omega^2} \left[\begin{array}{cc}
1+  \dfrac{6  \tau ^2 (\xi -\xi_\gamma )^2 h^4 }{ \bar{\gamma}^2\Omega^4+(\tau \chi^2+\xi_\gamma h^2)^2} \dfrac{\chi ^2}{\Omega ^2}  &-\dfrac{6\tau ^2  (\xi -\xi_\gamma )^2 h^2 \chi ^2}{\bar{\gamma}^2\Omega^4+(\tau \chi^2+\xi_\gamma h^2)^2}\dfrac{\chi h}{\Omega^2}  \\
-  \dfrac{6 \tau ^2 (\xi -\xi_\gamma )^2 h^2  \chi ^2 }{\bar{\gamma}^2\Omega^4+(\tau \chi^2+\xi_\gamma h^2)^2}  \dfrac{\chi h}{\Omega ^2}   & 1+\dfrac{6  \tau ^2  (\xi -\xi_\gamma )^2 \chi^4}{ \bar{\gamma}^2\Omega^4+(\tau \chi^2+\xi_\gamma h^2)^2}  \dfrac{h^2}{\Omega ^2} 
     \end{array}\right]\,.
 \end{equation} 
As argued in Section \ref{sec:HDSI}, for large $h$ values and at leading order in $\tau$, the  Gaussian curvature of this field manifold coincides with the one in the Palatini formulation,  cf. Eq.~\eqref{eq:kH}, making foreseeable that the contribution of the Holst term becomes negligible in that regime. To show this explicitly, we can diagonalize the kinetic sector of the theory by using the change of variables in Ref.~\cite{Rubio:2020zht},
\begin{equation} \label{Eq:Holst-diagonal}
\theta = \tau \,\frac{h^2+\chi^2}{\Omega^2}\,,
\hspace{20mm} \rho=\frac{1}{2\tau}\ln \left(\chi^2+h^2 \right)\,.
\end{equation}
This gives rise to an action of the form \eqref{SI-Lagrangian} with kinetic functions
\begin{equation}\label{Holst-pole}
K_\theta(\theta) = -\frac{1}{4\theta}\left(\frac{1}{\kappa_H \, \theta+c}-\frac{1}{\kappa_H(\theta-1)} \right)+F_\gamma(\theta)\,, \hspace{10mm}
K_\rho(\theta)= \theta \,, 
\end{equation}
and inflationary potential
\begin{equation}\label{eq:pottheta}
V(\theta)\simeq \frac{\lambda}{4\kappa_H^2}(1-\theta)^2\,.
\end{equation}
Here, we have safely neglected the contributions proportional to $\alpha,\beta\ll 1$, since these are not important at the field values explored during inflation. On top of that, we have explicitly isolated a \textit{non-singular} function of $\theta$ depending on $\bar{\gamma}$ and $\xi_\gamma$,
\begin{equation}\label{Eq:Fgamma}
 F_\gamma(\theta)= \frac{3  \left(\kappa_H+\left(1 +\frac{c}{\kappa_H}\right) \xi _{\gamma }\right)^2}{2 \,c\,  \kappa_H^2\left(1 +\frac{c}{\kappa_H}\right) \left[ \bar{\gamma }^2\kappa_H^2 \left(1 +\frac{c}{\kappa_H}\right) ^2 +\left(\left(1 +\frac{c}{\kappa_H}\right)  \xi _{\gamma }(\theta -1) +\kappa_H \left( \theta+\frac{c}{\kappa_H}\right)\right)^2\right]}\,,
\end{equation}
with
\begin{equation}\label{Eq:C}
c=\tau\left(1-\frac{\tau}{\xi}\right)\,.
\end{equation}
The structure in the first piece in Eq.~(\ref{Holst-pole}) is the same one obtained in Ref.~\cite{Rubio:2020zht}. In particular, the single pole at $\theta=1$ is a \textit{Minkowski} pole active only when the $\theta$ field reaches the minimum of the potential \eqref{eq:pottheta}, and therefore not relevant during inflation.
This implies that, when the main contribution to the kinetic term is coming from the poles at $\theta=0$ and $\theta=-c/\kappa_H$, we should expect a stretching of the canonically normalized inflaton field
\begin{equation}\label{eq:cantheta}
\tilde{\theta}\simeq \int^{\tilde{\theta}} \frac{d\theta}{\sqrt{-4\,  \theta
(\kappa_H \theta+c)}} \hspace{5mm}\rightarrow \hspace{5mm} 
\theta =
\begin{cases}
\exp\left(-2\sqrt{-\kappa_H}\,\tilde{\theta}\right)\,,\hspace{5mm}{\rm
for}\hspace{5mm} c=0 \ , \\ \frac{c}{-\kappa_H}\cosh
(\sqrt{-\kappa_H}\,\tilde{\theta}) \,, \hspace{5mm}{\rm for}\hspace{5mm} c\neq 
0\,,
\end{cases}
\end{equation}
leading  itself to the flattening of the potential~\cite{Galante:2014ifa,Karananas:2016kyt,Rubio:2018ogq,Artymowski:2016pjz}.
%%%%%%%%%%%%%%%%%%%%%%%%%%%%%%%%%%%%%%%%%%%%%%%%%%%%%%%%%%%%%%%%%%%%%%%%%%%%%%%%
\begin{figure}[ht]
    \begin{center}
        \fbox{\begin{minipage}[ht]{0.42\linewidth}
            \center{\includegraphics[width=\textwidth]{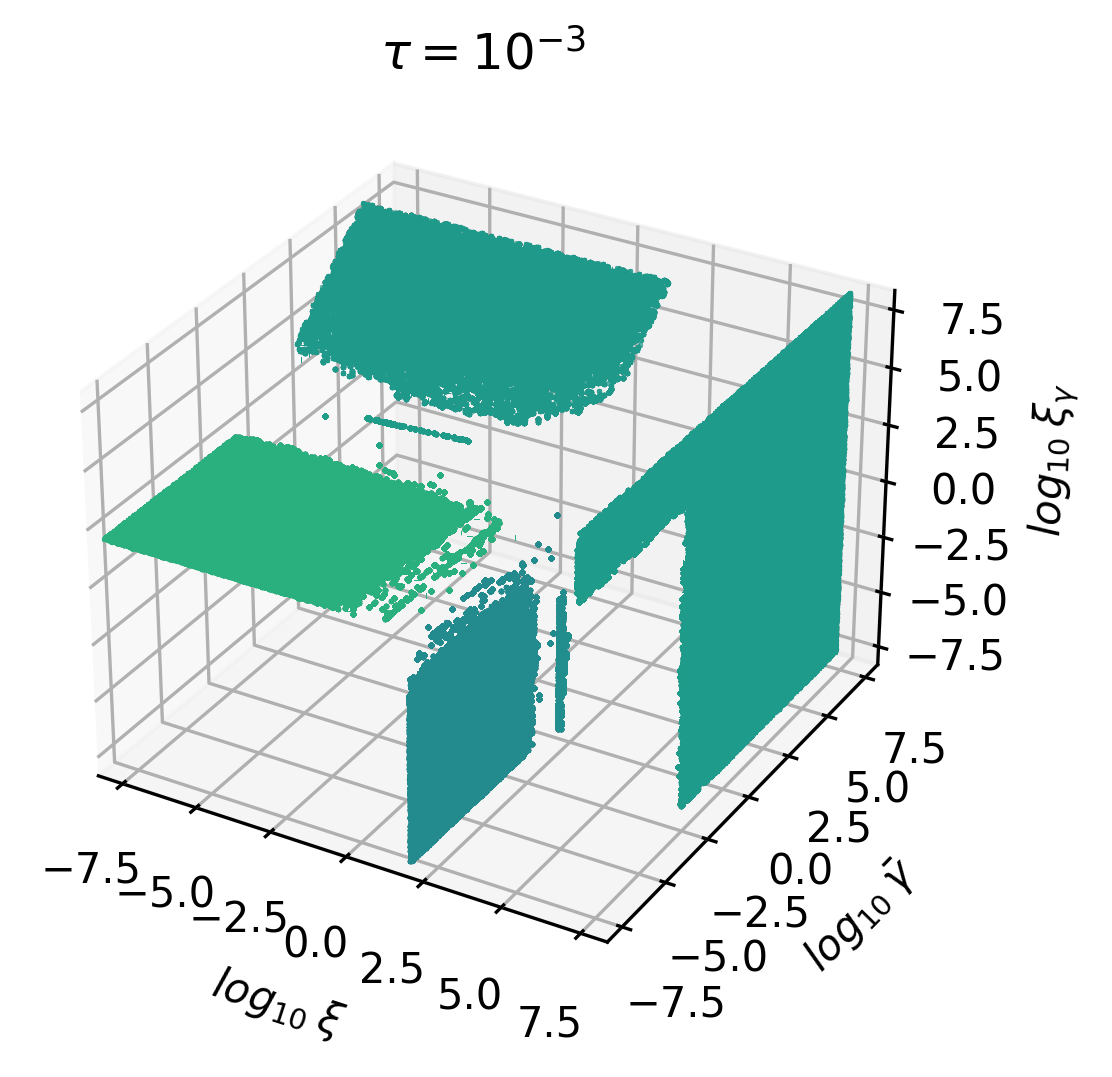} \\(a)}
        \end{minipage}}
        \fbox{\begin{minipage}[ht]{0.42\linewidth}
            \center{\includegraphics[width=\textwidth]{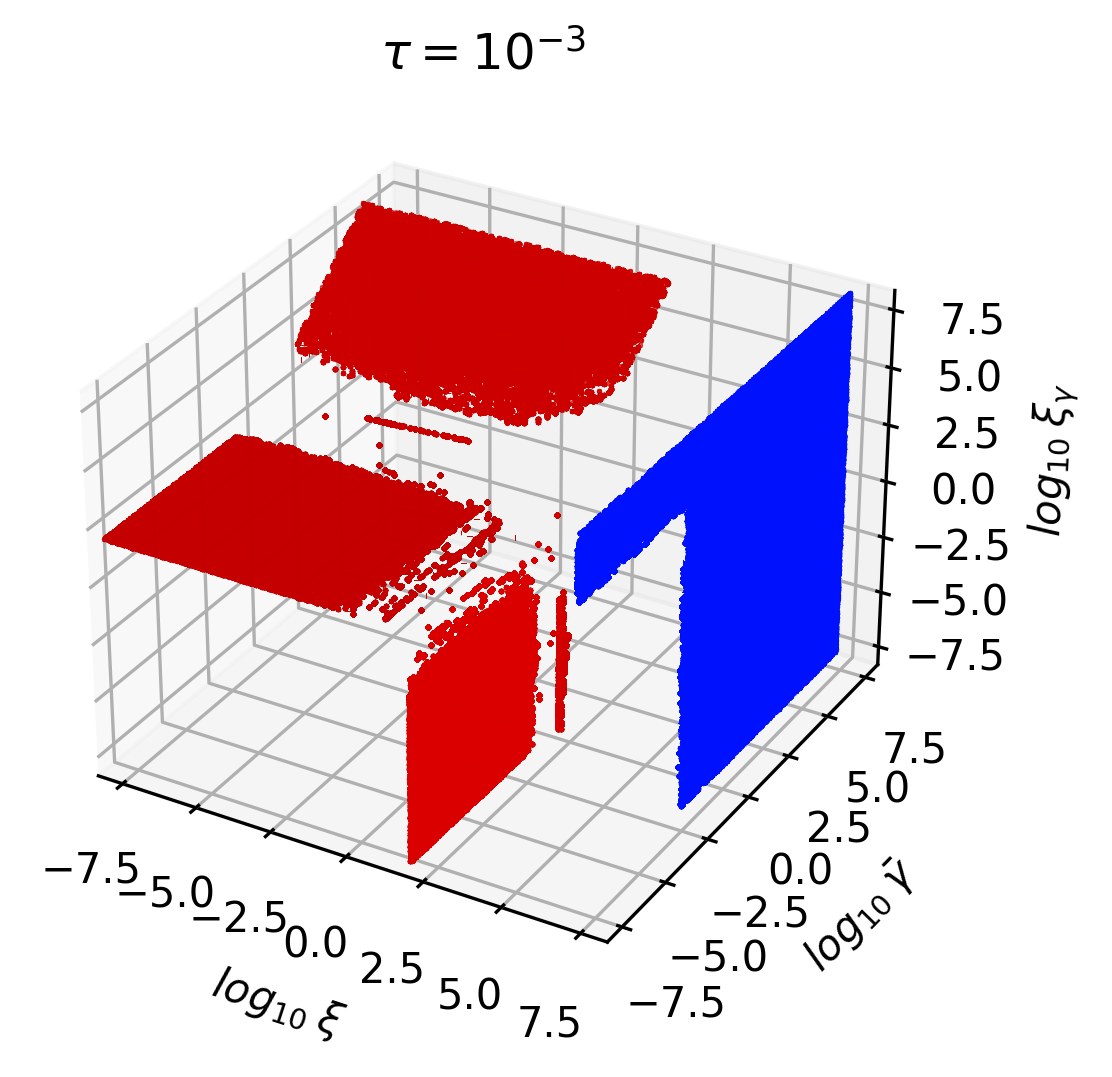} \\(b)}
        \end{minipage}}
        \fbox{\begin{minipage}[ht]{0.42\linewidth}
            \center{\includegraphics[width=\textwidth]{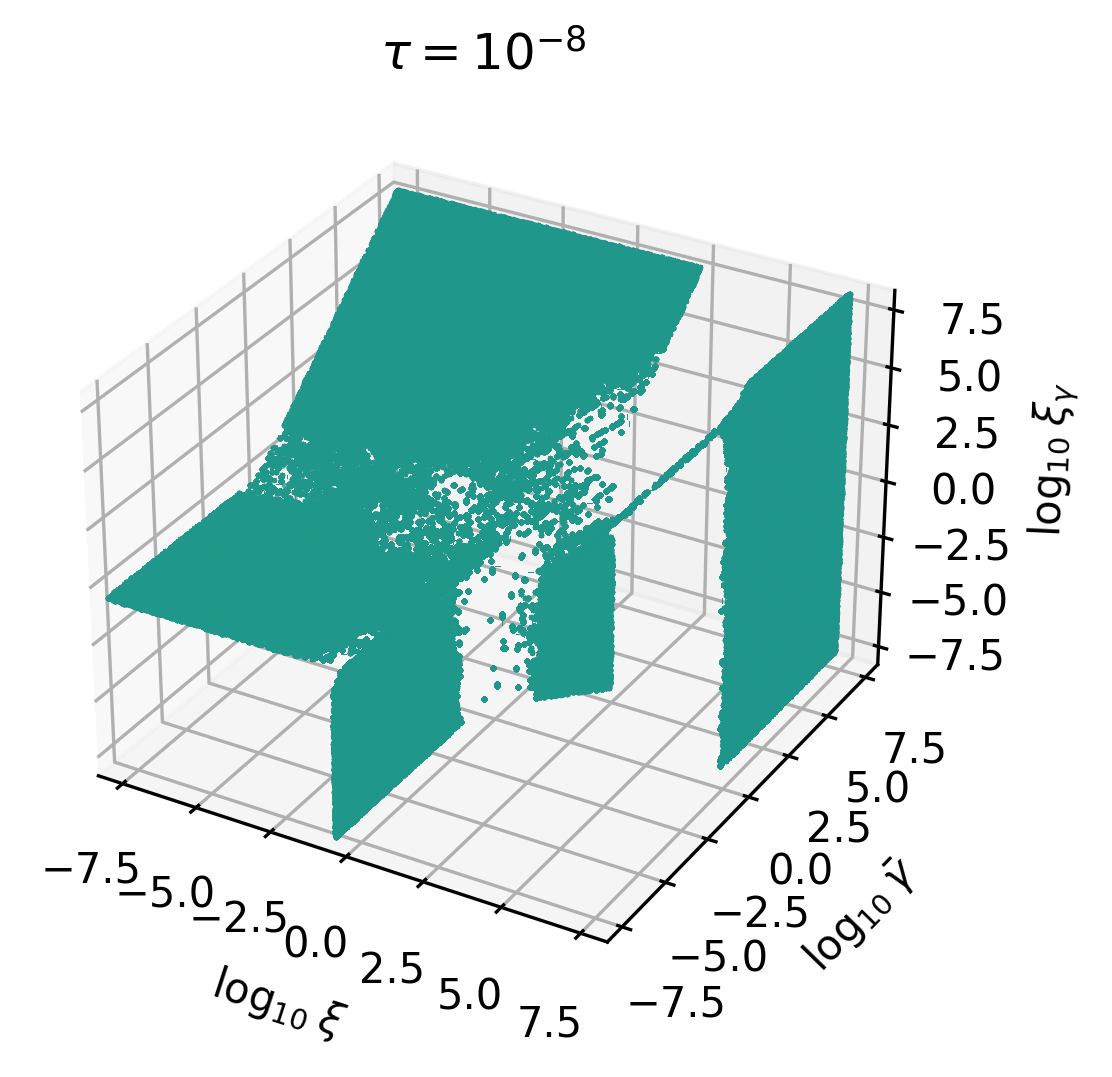} \\(c)}
             
        \end{minipage}}
               \fbox{\begin{minipage}[ht]{0.42\linewidth}
            \center{\includegraphics[width=\textwidth]{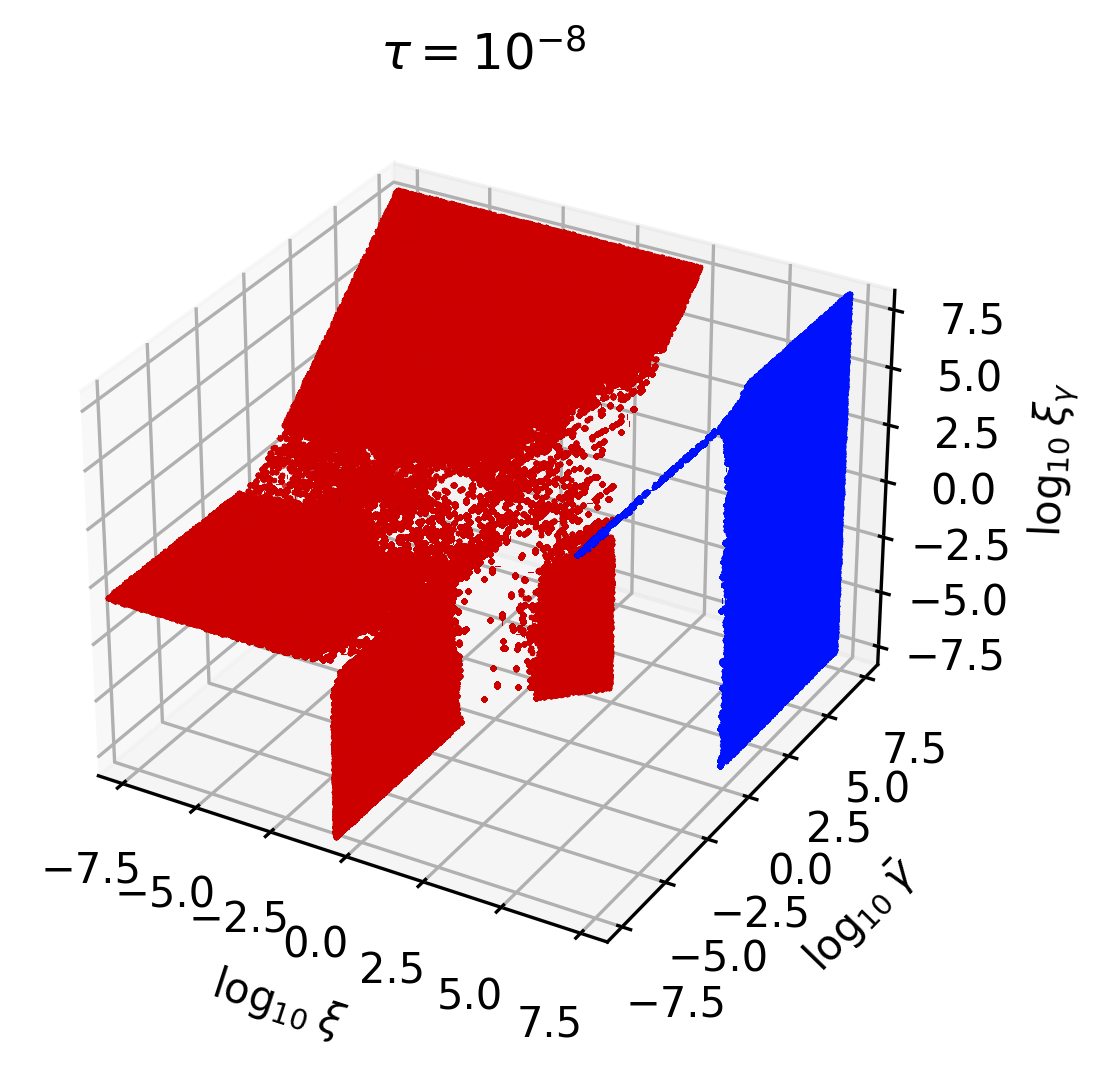} \\(d)}
             
        \end{minipage}}
                          \begin{minipage}[ht]{0.42\linewidth}
            \center{\includegraphics[width=\textwidth]{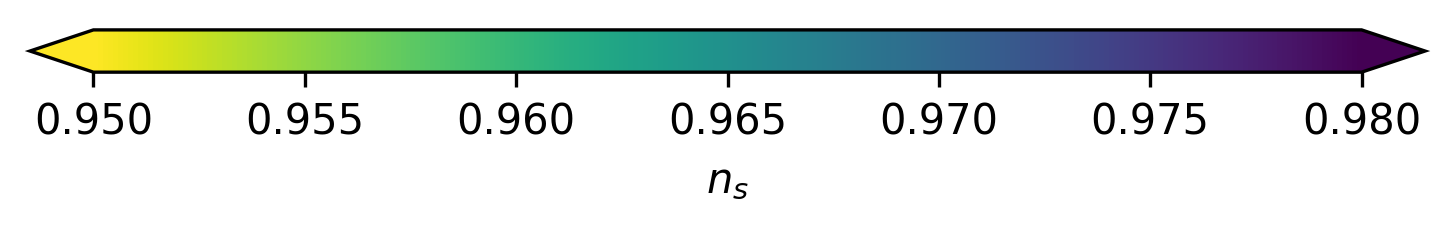}}
             \end{minipage}
                \begin{minipage}[ht]{0.42\linewidth}
            \center{\includegraphics[width=\textwidth]{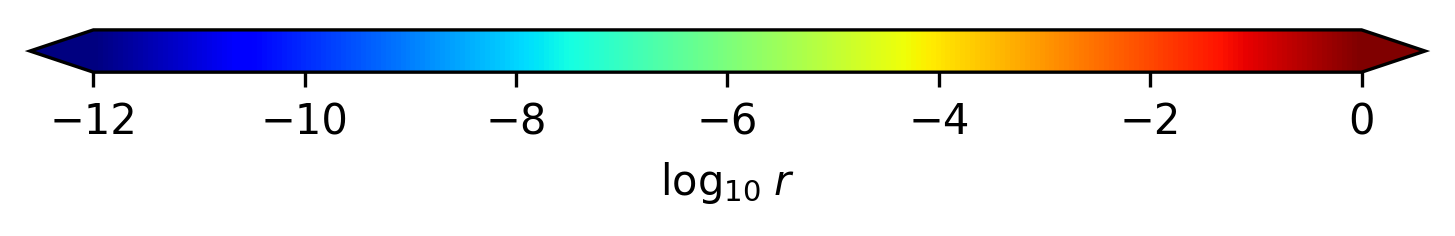}}
             \end{minipage}
    \caption{Numerical results for the Holst-like inflation case at fixed $\tau$ values. The left and right panels display respectively the results for the spectral tilt and tensor-to-scalar ratio, with the different colors corresponding to different numerical values, as indicated in the figure. The Palatini-like region, corresponding to the surface at constant $\xi \sim {\cal O}(10^7)$, is the only one not exceeding the current bounds on the tensor-to-scalar ratio, being all other cases ruled out. For low values of $\bar{\gamma}$, the surface becomes discontinuous due to the absence of points satisfying the constraint $N=55$ on the duration of inflation.}
    \label{Fig:H-plots}
    \end{center}
\end{figure}
%%%%%%%%%%%%%%%%%%%%%%%%%%%%%%%%%%%%%%%%%%%%%%%%%%%%%%%%%%%%%%%%%%%%%%%%%%%%%%%%
%%%%%%%%%%%%%%%%%%%%%%%%%%%%%%%%%%%%%%%%%%%%%%%%%%%%%%%%%%%%%%%%%%%%%%%%%%%%%%%%
\begin{figure}
    \begin{center}
        \fbox{\begin{minipage}[ht]{0.42\linewidth}
            \center{\includegraphics[width=\textwidth]{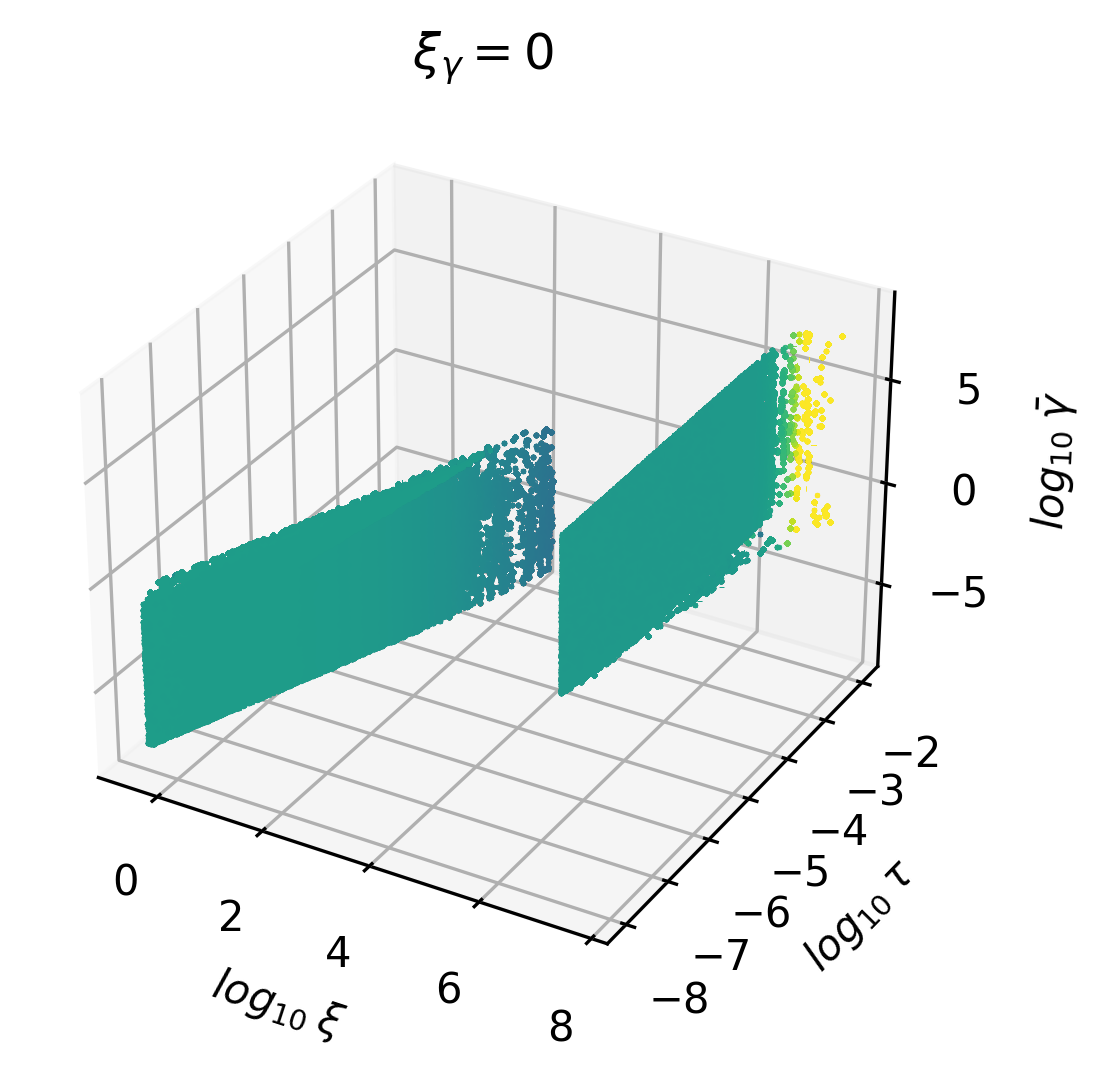} \\(a)}
        \end{minipage}}
        \fbox{\begin{minipage}[ht]{0.42\linewidth}
            \center{\includegraphics[width=\textwidth]{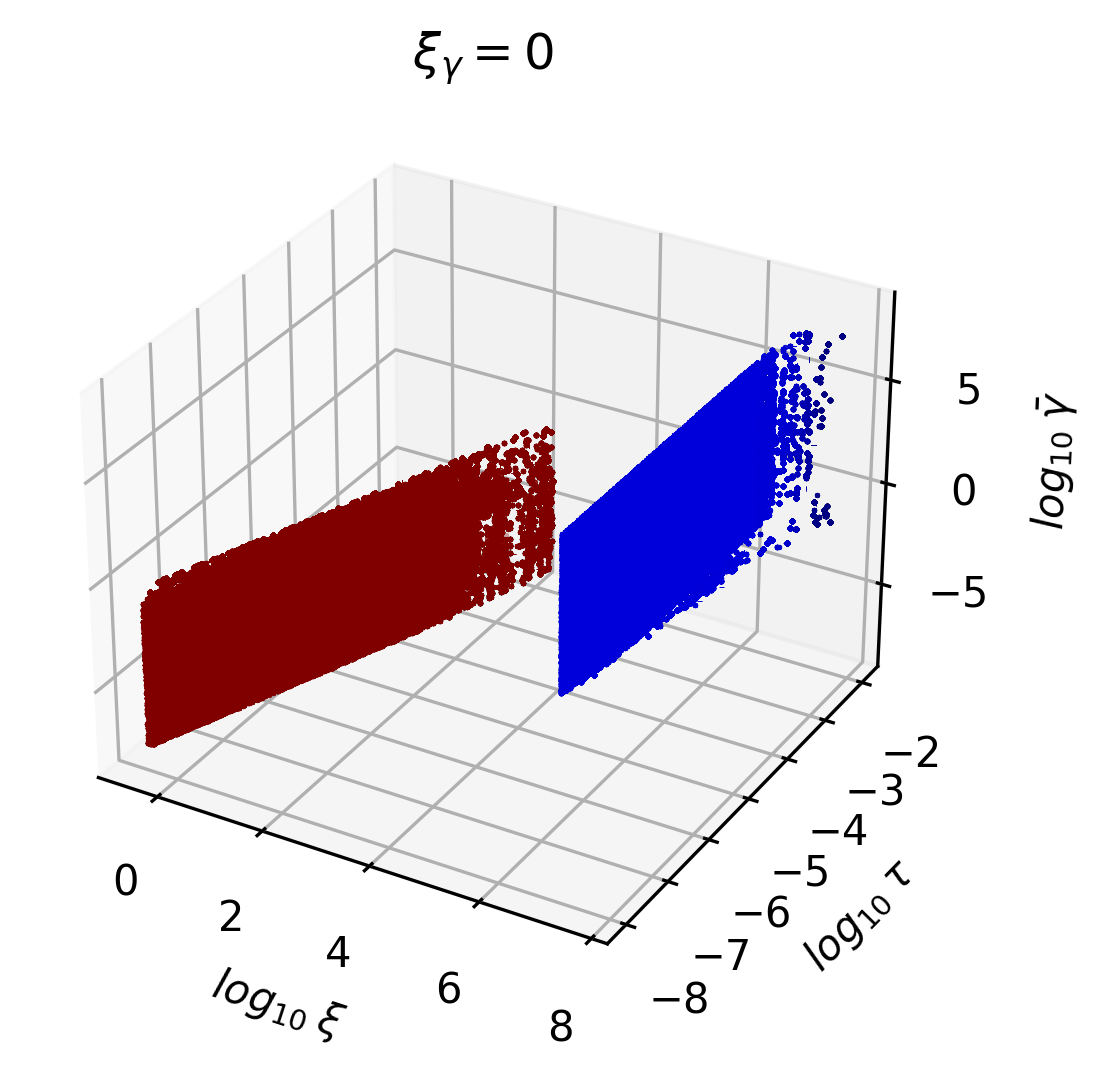} \\(b)}
        \end{minipage}}
                   \begin{minipage}[ht]{0.42\linewidth}
            \center{\includegraphics[width=\textwidth]{Cbar_NS-std.png}}
             \end{minipage}
                \begin{minipage}[ht]{0.42\linewidth}
            \center{\includegraphics[width=\textwidth]{Cbar_R-std.png}}
             \end{minipage}
        \caption{Numerical results for the Holst-like inflation case at vanishing $\xi_\gamma$. The left and right panels display respectively the results for the spectral tilt and the tensor-to-scalar ratio, with the different colors corresponding to different numerical values, as indicated in the figure. From panel (a), we see that the spectral tilt decreases with $\tau$, leading eventually to a violation of the slow-roll conditions. Note also that, in the regions disfavoured by the observational constraints on the tensor-to-scalar ratio, the values of $\xi$ depend on $\tau$.}
    \label{fig:Holst-SRB}
    \end{center}
\end{figure}
%%%%%%%%%%%%%%%%%%%%%%%%%%%%%%%%%%%%%%%%%%%%%%%%%%%%%%%%%%%%%%%%%%%%%%%%%%%%%%%%
 In this limit, and for non-vanishing $c\ll -\kappa_H$, one can easily determine the following approximate expressions for the spectral tilt and the tensor-to-scalar ratio \cite{Casas:2018fum,Almeida:2018oid,Rubio:2020zht},
\begin{equation}\label{ns-r-polestotal}
    n_s\simeq  1-8\, c \coth (4 c N)\,, \hspace{15mm} r\simeq \frac{32 c^2}{-\kappa_H}\csch^2(4 c N) \,.
\end{equation}
These compact formulas interpolate between the well-known Palatini-like attractor
\begin{equation}\label{ns-r-poles}
    n_s\simeq  1-\frac{2}{N}\,, \hspace{15mm} r\simeq \frac{2}{-\kappa_H N^2}\,,
\end{equation}
at $4 c N\ll 1$ and an asymptotic behaviour
\begin{equation}\label{ns-r-singlepole}
    n_s\simeq  1-8c\,, \hspace{15mm} r\simeq 0\,,
\end{equation}
at $4 c N\gg 1$. This is confirmed by the numerical analysis in Figs.~\ref{Fig:H-plots} and \ref{fig:Holst-SRB}, where we report the values of the spectral tilt and the tensor-to-scalar ratio for different three-dimensional slices in parameter space. In particular, by fixing the value of $\tau$ in Fig.~\ref{Fig:H-plots}, we can clearly distinguish the above Palatini-like limit in the surfaces of constant $\xi \sim \mathcal{O}(10^7)$. As expected, the phenomenology in this regime does not depend on the value of $\xi_\gamma$, a direct consequence of the fact that inflation is taking place in the vicinity of the poles in Eq.~\eqref{Holst-pole}, where the regular function $F_\gamma(\theta)$ plays a completely subdominant role.  Moreover, in accordance with Eq.~\eqref{ns-r-poles}, the spectral tilt and the tensor-to-scalar ratio are of the order $n_s\simeq 0.964$ and $r\sim\mathcal{O}(10^{-10})$.  The mild dependence of the dilaton coupling $\tau$ becomes apparent in Fig.~\ref{fig:Holst-SRB}, where we display the same observables for vanishing $\xi_\gamma$. The trend presented there is in line with the asymptotic behaviour in Eq.~\eqref{ns-r-singlepole}, where the value of $n_s$ decreases as $c$ increases, leading eventually to the breaking of the slow-roll conditions. Indeed, as shown in detail in Fig.~\ref{Fig:NS-C}, the agreement with the analytical prediction \eqref{ns-r-polestotal} is excellent.  Note, also that, as illustrated in Fig.~\ref{fig:runningsH}, the running of the spectral tilt and the running of the running are small and negative for all the sampled parameters, in accordance with the 68\%\;\text{C.L.} constraints on these quantities, $\alpha_s = 0.002 \pm 0.010$, $\beta_s = 0.010 \pm 0.013$ \cite{Planck:2018jri}.
%%%%%%%%%%%%%%%%%%%%%%%%%%%%%%%%%%%%%%%%%%%%%%%%%%%%%%%%%%%%%%%%%%%%%%%%%%%%%%%%
%%%%%%%%%%%%%%%%%%%%%%%%%%%%%%%%%%%%%%%%%%%%%%%%%%%%%%%%%%%%%%%%%%%%%%%%%%%%%%%%
%%%%%%%%%%%%%%%%%%%%%%%%%%%%%%%%%%%%%%%%%%%%%%%%%%%%%%%%%%%%%%%%%%%%%%%%%%%%%%%%
%%%%%%%%%%%%%%%%%%%%%%%%%%%%%%%%%%%%%%%%%%%%%%%%%%%%%%%%%%%%%%%%%%%%%%%%%%%%%%%%

For lower values of $\xi$ (or smaller $\vert \kappa_H\vert$), we find regions where the function $F_\gamma(\theta)$ becomes relevant. However, the tensor-to-scalar ratio in these surfaces increases significantly, exceeding the current observational bounds on this quantity and ruling out all inflationary regimes but the Palatini one. Our findings are in line with the single-field scenario~\cite{Langvik:2020nrs,Shaposhnikov:2020gts}, up to differences associated with a degeneracy between $\tau$ and $\xi$ that becomes boosted at $\bar\gamma\ll 1$, cf.~Eqs.~\eqref{Holst-action} and \eqref{Eq:Fgamma}. In particular, a direct comparison of the plots in Fig.~\ref{Fig:H-plots} with their HI counterparts shows that in this limit the allowed values for $\xi$ are shifted as a function of $\tau$. 

%%%%%%%%%%%%%%%%%%%%%%%%%%%%%%%%%%%%%%%%%%%%%%%%%%%%%%%%%%%%%%%%%%%%%%%%%%%%%%%%
\subsection{Nieh-Yan inflation}\label{sec:NY}
%%%%%%%%%%%%%%%%%%%%%%%%%%%%%%%%%%%%%%%%%%%%%%%%%%%%%%%%%%%%%%%%%%%%%%%%%%%%%%%%

\begin{figure}
    \centering
    \includegraphics[scale=0.8]{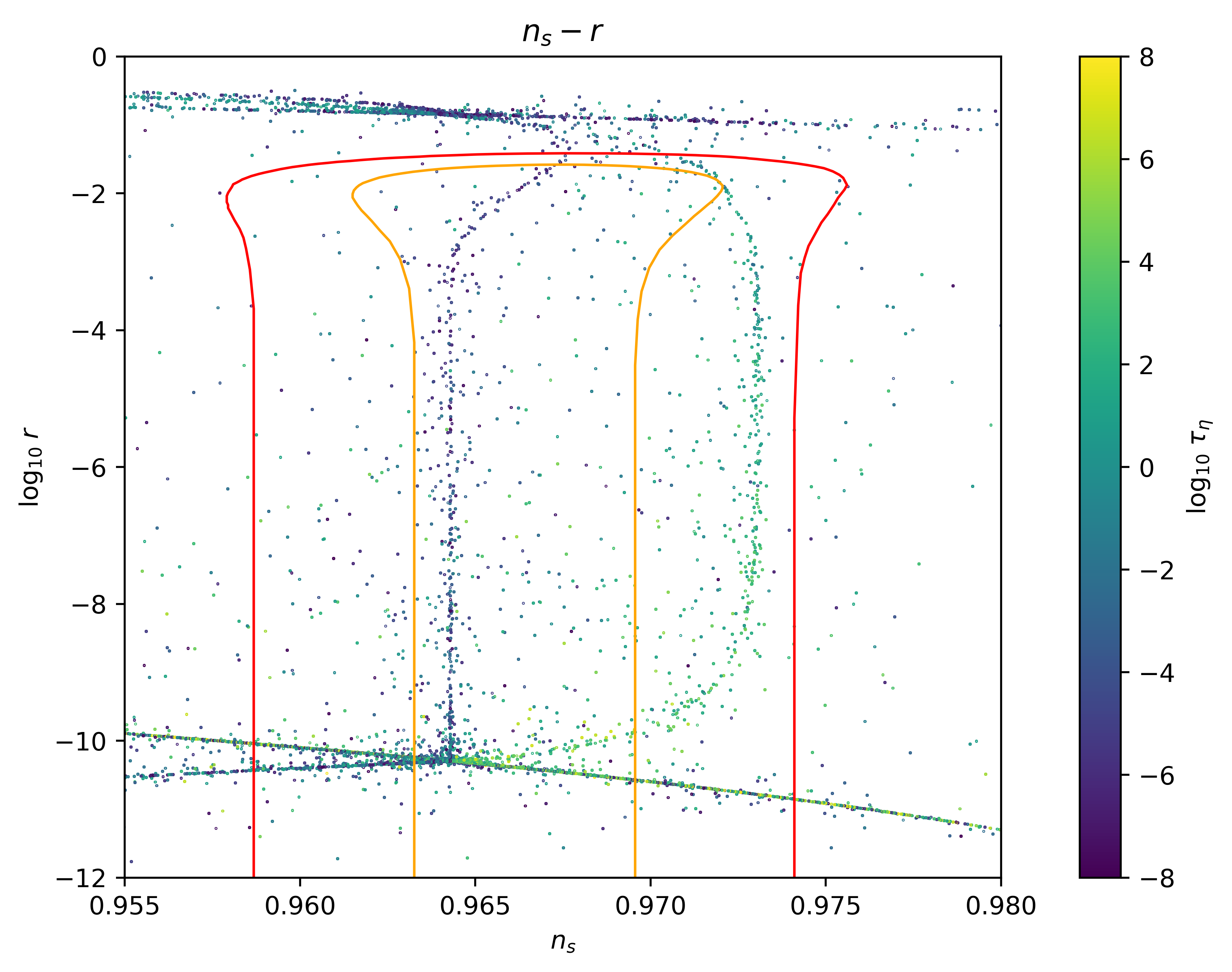}
    \caption{Numerical result for the MCMC scan of the full parameter space of Nieh-Yan-like inflation in the $n_s-r$ plane, together with the recently updated constraints (yellow and red curves) from the \textit{BICEP/Keck} collaboration~\cite{BICEP:2021xfz}, at 68\%\; and 95\%\;\text{C.L.}. The color gradient represents the value of $\tau_\eta$ in a logarithmic scale. The two attractor behaviours correspond respectively to the quadratic (left) and cubic (right) pole dominance in Eq.~\eqref{NY-Ks}.}
    \label{fig:ns-r}
\end{figure}

The second case we consider is Nieh-Yan-like inflation. This is obtained by setting $\bar{\gamma}\rightarrow \infty,\;\xi_\gamma=0$ in Eq.~\eqref{K-matrix},
 \begin{equation}\label{NY-kin}
  K_{ab}= \frac{1}{\Omega^2} \left[\begin{array}{cc}
 1+   6  \tau_{\eta}^2  \dfrac{\chi^2 }{\Omega^2}   &  6\tau_\eta \xi_\eta  \dfrac{ h \chi }{\Omega^2} \\
6\tau_\eta \xi_\eta   \dfrac{h \chi}{\Omega^2}    &1+   6  \xi_{\eta}^2 \dfrac{h^2}{\Omega^2} 
     \end{array}\right]\,.
 \end{equation} 
We notice immediately that for $\xi_\eta=\xi$, $\tau_\eta=\tau$ one recovers the metric case, while the limit $\xi_\eta=\tau_\eta=0$ corresponds to the Palatini one. 
As we did for Holst-like inflation, we can easily reduce the kinetic sector of the theory to a diagonal form by switching to a set of variables
\begin{equation}
\theta = \tau\, \frac{h^2+\chi^2}{\Omega^2}\,,\hspace{20mm} \rho=\rho(h,\chi)\,,
\end{equation}
with the functional form of $\rho(h,\chi)$ dictated by the conservation of the dilatation current (cf.~Appendix~\ref{appendix:diagonalization} for details). In terms of these quantities, the kinetic functions and the inflationary potential in Eq.~\eqref{SI-Lagrangian} become respectively
\begin{eqnarray}\label{NY-Ks}
 K_\theta(\theta)&=&
 -\frac{1}{4\theta} \left[\frac{1}{ \kappa_H\,\theta+c}-\frac{1}{\kappa_H (\theta-1 )}-\frac{a^2}{(\theta-\theta_+)(\theta-\theta_-)}\right] \,,\\
K_\rho(\theta)&=& \frac{1}{2\, \tau \sigma\kappa_H^2} (\theta-\theta_+)(\theta-\theta_-)\,,
\label{eq:Krho}\\
V(\theta) &\simeq&\frac{\lambda}{4\kappa_H^2} (1-\theta)^2\,,\label{NY-Ks_pot}
\end{eqnarray}
where we have neglected again the contributions proportional to $\alpha,\beta\ll 1$ and defined 
\begin{equation} \label{adef}
\theta_{\pm}=T\mp \Delta\,, \hspace{15mm} a^2=12\, \sigma (\tau_\eta-\xi_\eta)^2\,,
\end{equation}
with
\begin{equation}
T=\sigma\left[\kappa_{NY} \left(\xi +6 \xi_\eta ^2\right)+\kappa_{\rm min} \left(\tau +6 \tau_\eta^2\right)\right]\,, \hspace{10mm} \Delta=\sigma\,\kappa_H (2 T-\kappa_H^2)^{1/2}\,,
\end{equation}
and
\begin{equation}
     \kappa_{\rm min}=-\frac{\tau(\tau-\xi)+12 \tau_\eta(\xi_\eta \tau - \xi \tau_\eta)}{\tau+6 \tau_\eta^2}\,, \hspace{10mm} \sigma=  \frac{\tau}{12\left( \xi_\eta \,\tau -\xi \tau_\eta\right)^2}\,.
\end{equation}
The form of Eq.~\eqref{NY-Ks} is particularly enlightening. While the first two terms~in this expression coincide exactly with those in the Holst-like action \eqref{Holst-pole}, the third one is no longer regular at all field values, but rather displays an explicit pole structure. This has important consequences. On the one hand, for parameter values leading to non-degenerate pole locations, $\theta_+\neq \theta_-$, the contribution of the third piece may become approximately quadratic and potentially relevant at $\theta\sim - c/\kappa_{H}$,  modifying with it the overall residue of the pole and the associated attractor solution in Eq.~\eqref{ns-r-poles}.  On the other hand, for an almost vanishing $\Delta$ leading to degenerate roots $\theta_+\simeq \theta_-\simeq T \ll 1$, the pole structure become approximately cubic. This gives rise to a new attractor solution where the spectral tilt and the tensor-to-scalar ratio approach the values \cite{Galante:2014ifa,Karamitsos:2021mtb}
\begin{equation}\label{eq:triple}
n_s\simeq 1-\frac{3}{2N}\,, \hspace{20mm} r\simeq \frac{2\,a}{N^{1/3}} \,.
\end{equation}
The result of sampling the complete parameter space of this model is illustrated in Fig.~\ref{fig:ns-r}, where we associate the values of the spectral tilt and the tensor-to-scalar ratio to that of the dilaton coupling $\tau_\eta$ for all admitted MCMC points. As anticipated above, we can clearly distinguish two visible attractor solutions correlated with the value of $\tau_\eta$, in qualitative agreement with the pole structure in Eq.~\eqref{NY-Ks}. To quantify this further, we analyze in what follows some particular slices of the parameter space. 

%%%%%%%%%%%%%%%%%%%%%%%%%%%%%%%%%%%%%%%%%%%%%%%%%%%%%%%%%%%%%%%%%%%%%%%%%%%%%%%%
%%%%%%%%%%%%%%%%%%%%%%%%%%%%%%%%%%%%%%%%%%%%%%%%%%%%%%%%%%%%%%%%%%%%%%%%%%%%%%%%
\begin{figure}[ht]
    \begin{center}
        \fbox{\begin{minipage}[h]{0.42\linewidth}
            \center{\includegraphics[width=\textwidth]{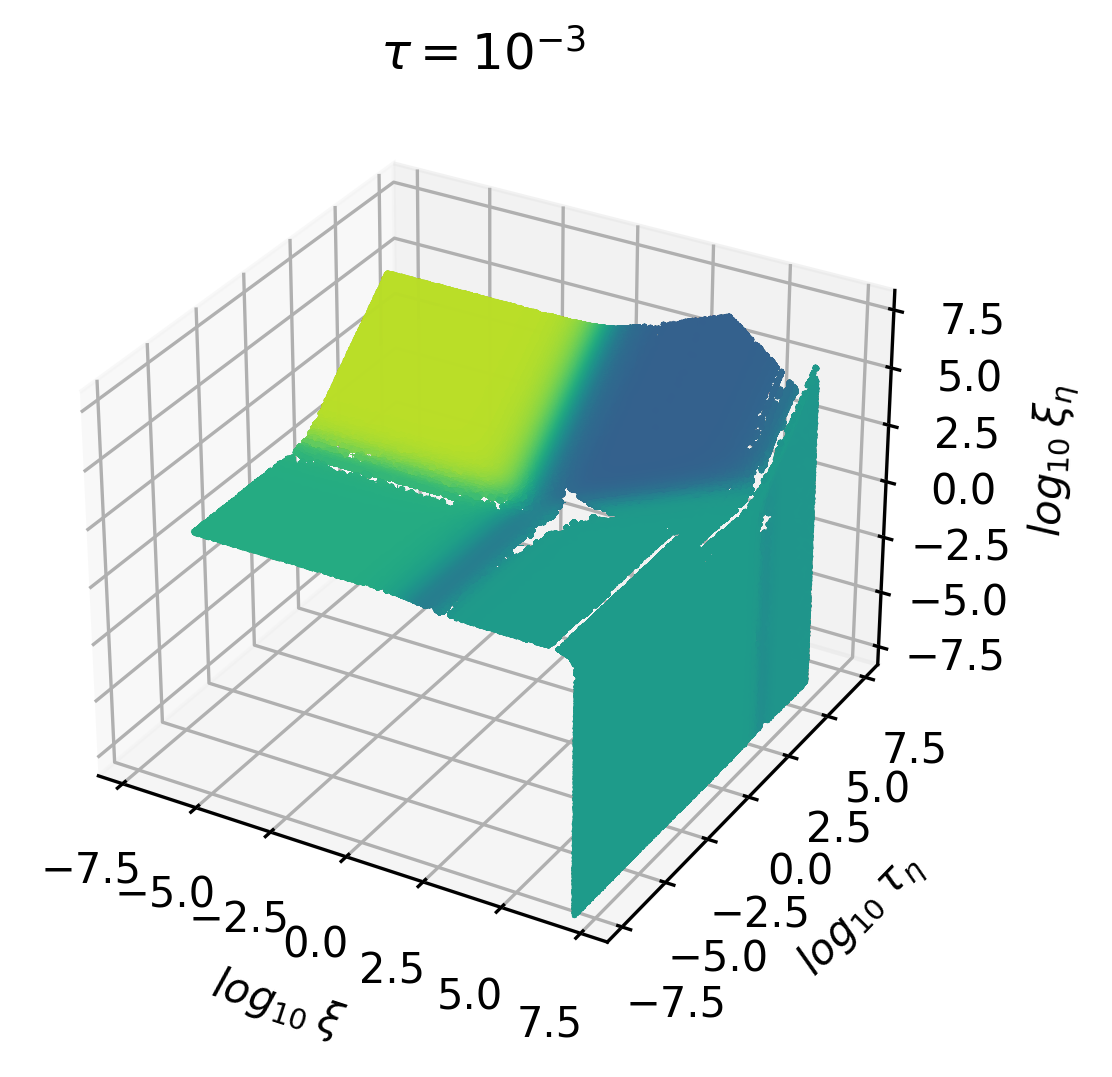}} \\(a)
        \end{minipage}}
        \fbox{\begin{minipage}[h]{0.42\linewidth}
           \center{\includegraphics[width=\textwidth]{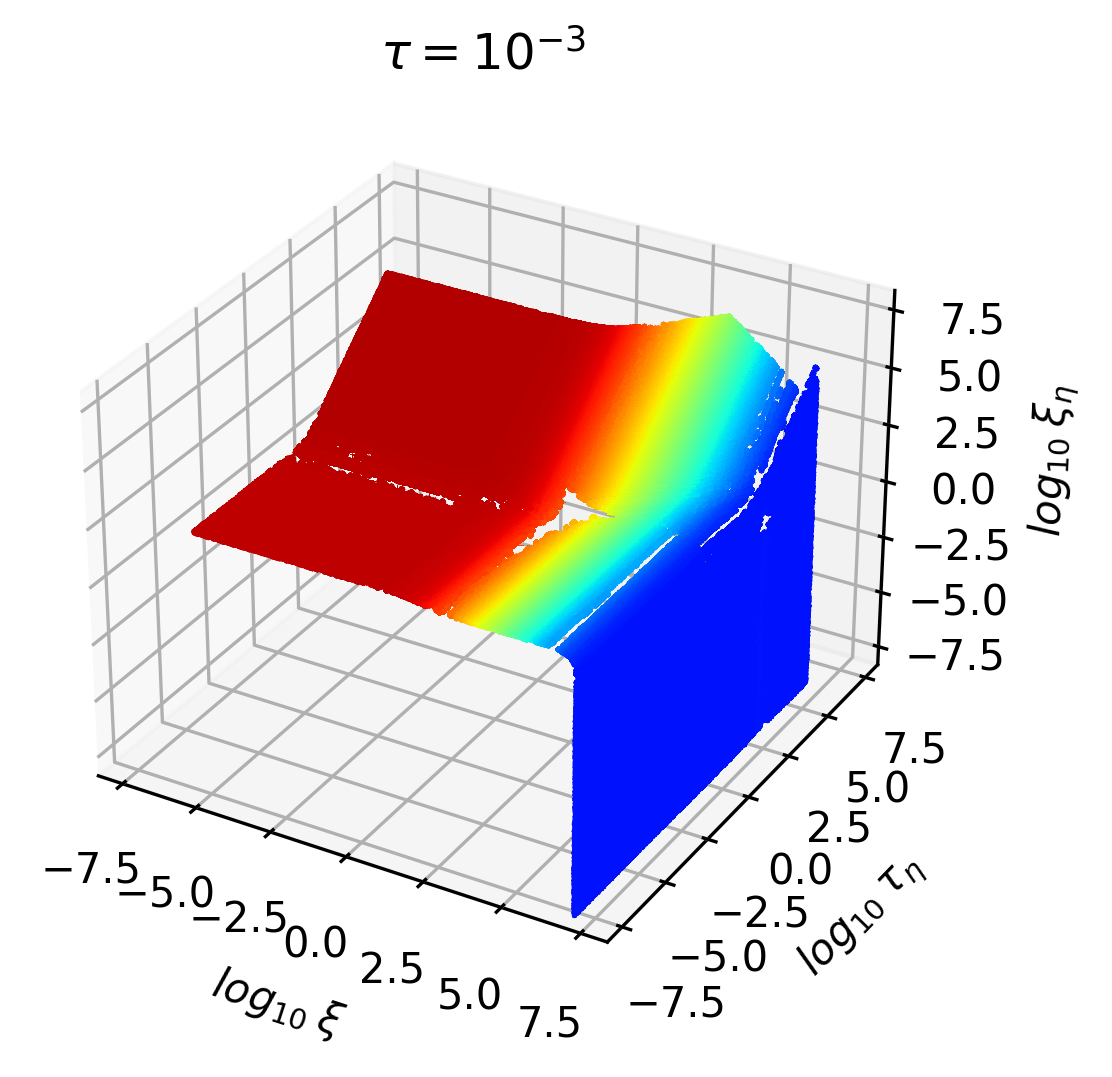} \\(b)}
        \end{minipage}}
        \fbox{\begin{minipage}[ht]{0.42\linewidth}
            \center{\includegraphics[width=\textwidth]{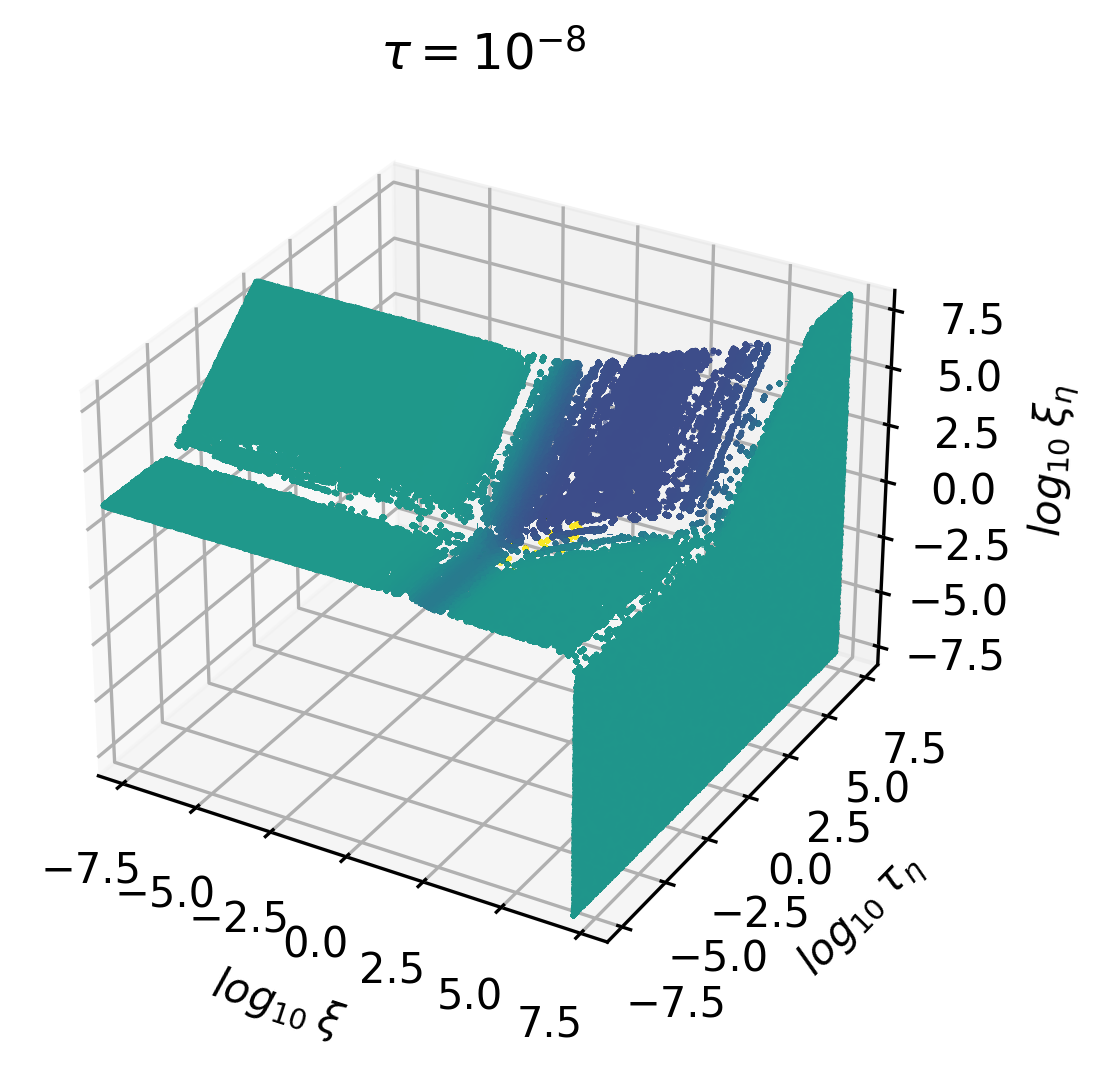} \\(c)}
             
        \end{minipage}}
               \fbox{\begin{minipage}[ht]{0.42\linewidth}
            \center{\includegraphics[width=\textwidth]{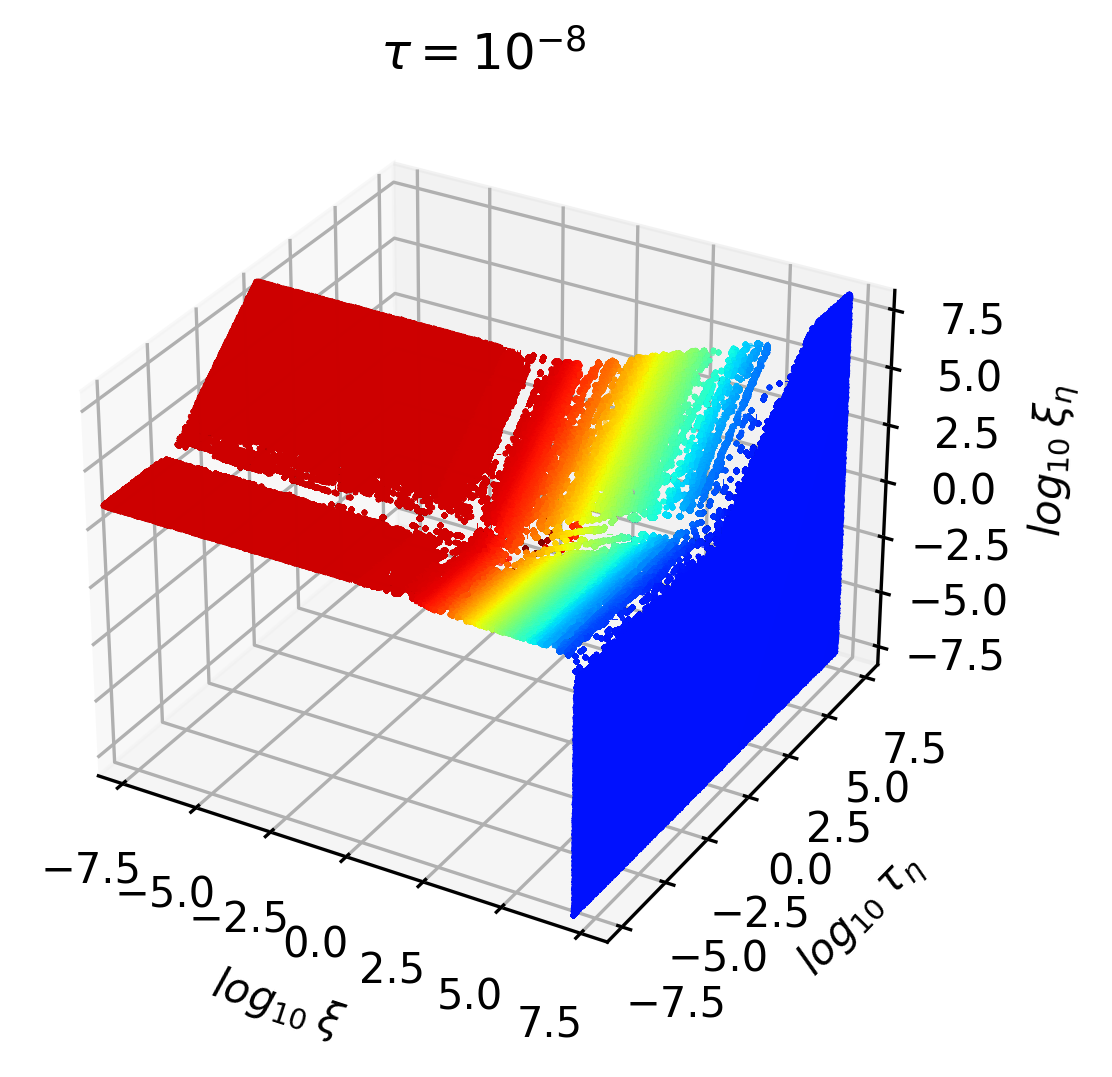} \\(d)}
             \end{minipage}}
                  \begin{minipage}[ht]{0.42\linewidth}
            \center{\includegraphics[width=\textwidth]{Cbar_NS-std.png}}
             \end{minipage}
                \begin{minipage}[ht]{0.42\linewidth}
            \center{\includegraphics[width=\textwidth]{Cbar_R-std.png}}
             \end{minipage}
    \caption{Numerical results for the Nieh-Yan-like inflation case at fixed $\tau$ values. The left and right panels display respectively the results for the spectral tilt and tensor-to-scalar ratio, with the different colors corresponding to different numerical values, as indicated in the figure.
    For small values of $\tau$ and $\tau_\eta$ the single-field HI limit ~\cite{Langvik:2020nrs,Shaposhnikov:2020gts} is always recovered.}
    \label{fig:NY-plots}
    \end{center}
\end{figure}
%%%%%%%%%%%%%%%%%%%%%%%%%%%%%%%%%%%%%%%%%%%%%%%%%%%%%%%%%%%%%%%%%%%%%%%%%%%%%%%%
\begin{figure}
    \begin{center}
        \fbox{\begin{minipage}[ht]{0.42\linewidth}
            \center{\includegraphics[width=\textwidth]{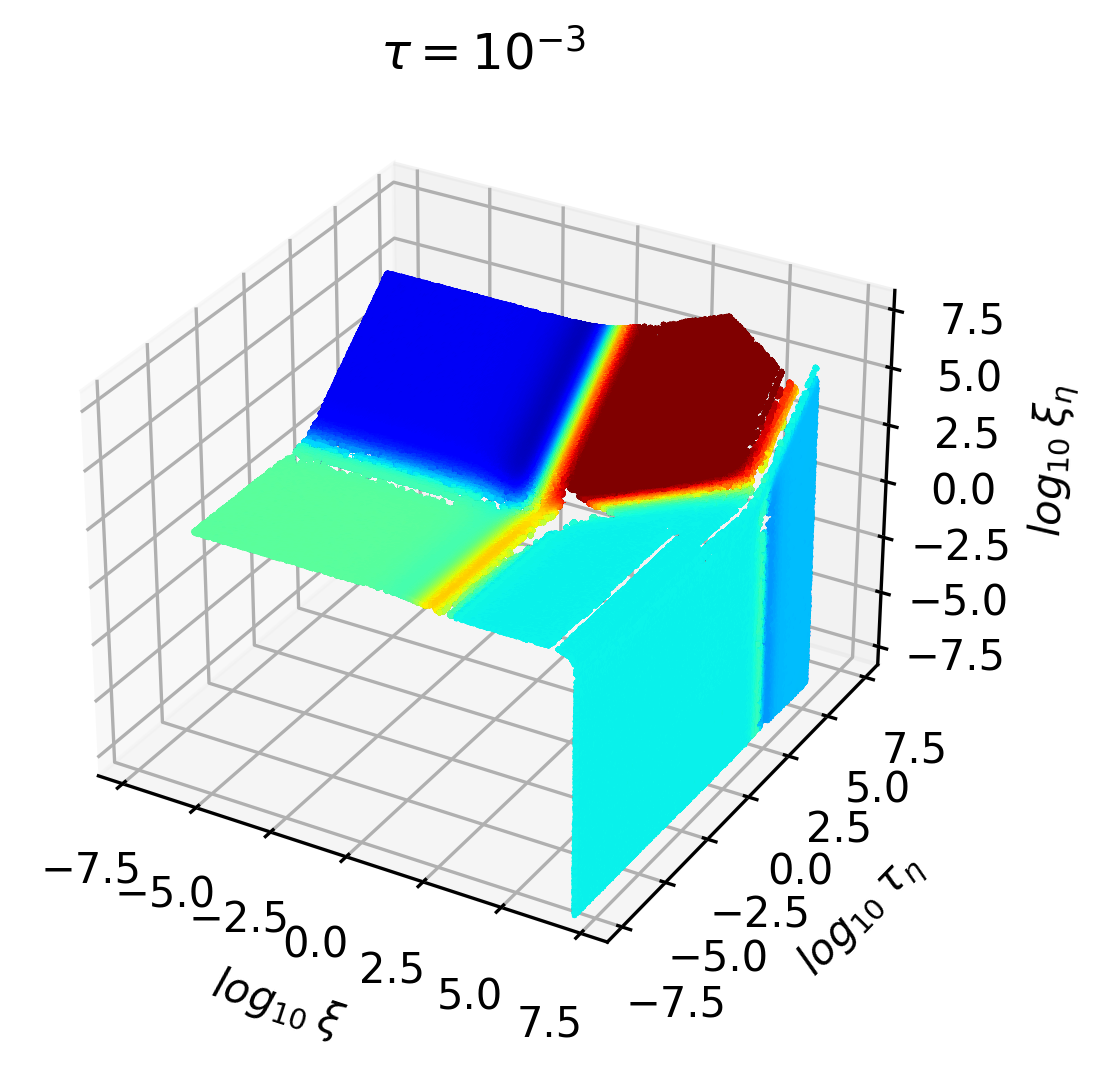} \\(a)}
        \end{minipage}}
        \fbox{\begin{minipage}[ht]{0.42\linewidth}
            \center{\includegraphics[width=\textwidth]{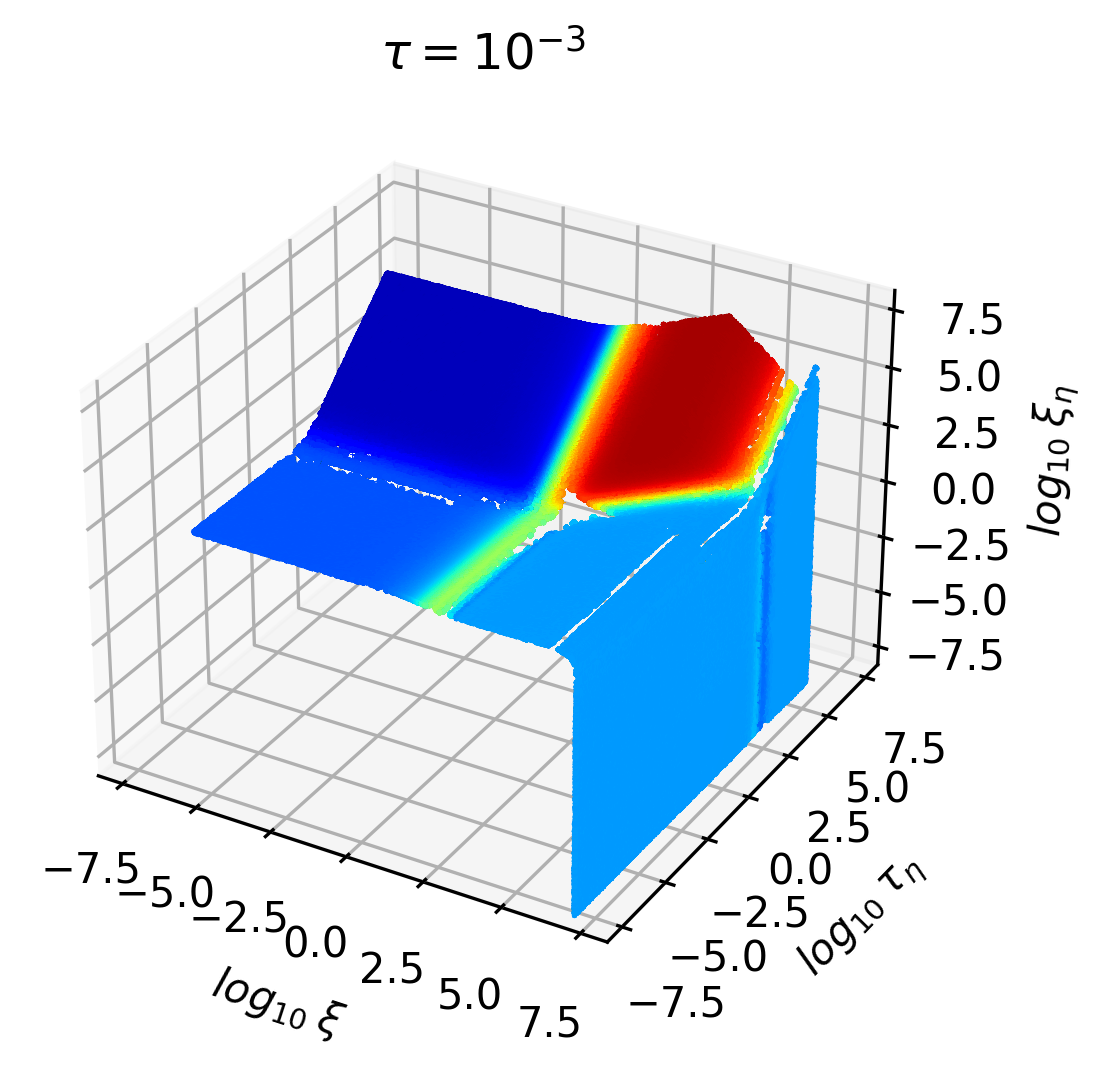} \\(b)}
        \end{minipage}}
                          \begin{minipage}[ht]{0.42\linewidth}
            \center{\includegraphics[width=\textwidth]{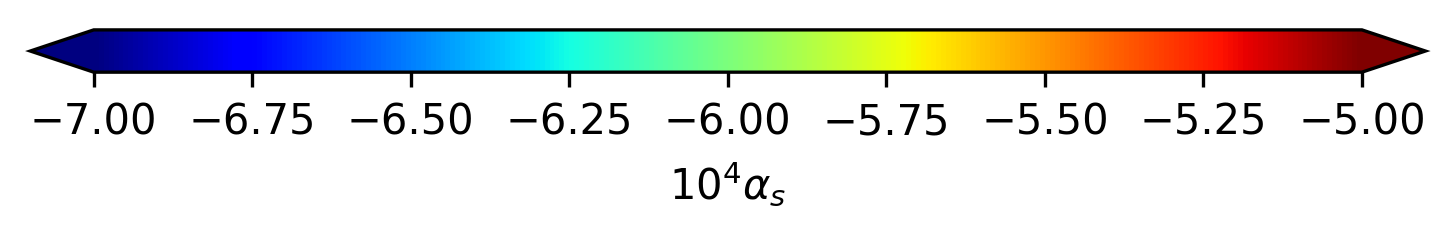}}
             \end{minipage}
                \begin{minipage}[ht]{0.42\linewidth}
            \center{\includegraphics[width=\textwidth]{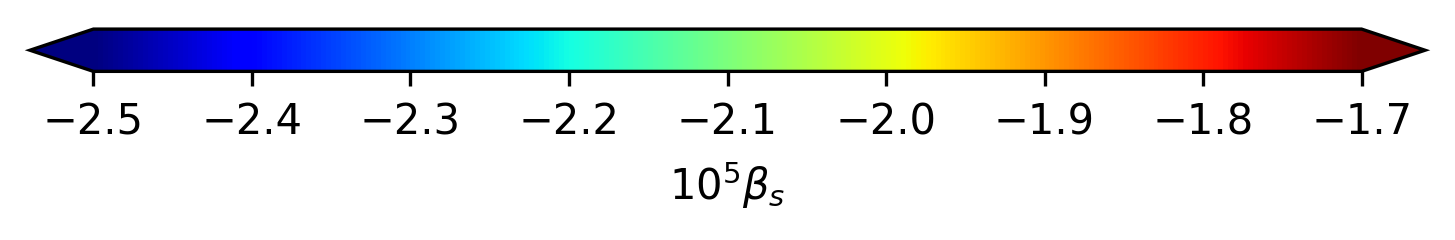}}
             \end{minipage}
    \caption{Numerical results for the running of the scalar spectral tilt  (left panels) and the running of the running (right panels) for a fixed value of the dilaton coupling $\tau=10^{-3}$.} \label{fig:runnings}
    \end{center} 
\end{figure}
\begin{figure}[ht]
    \begin{center}
        \fbox{\begin{minipage}[ht]{0.42\linewidth}
            \center{\includegraphics[width=\textwidth]{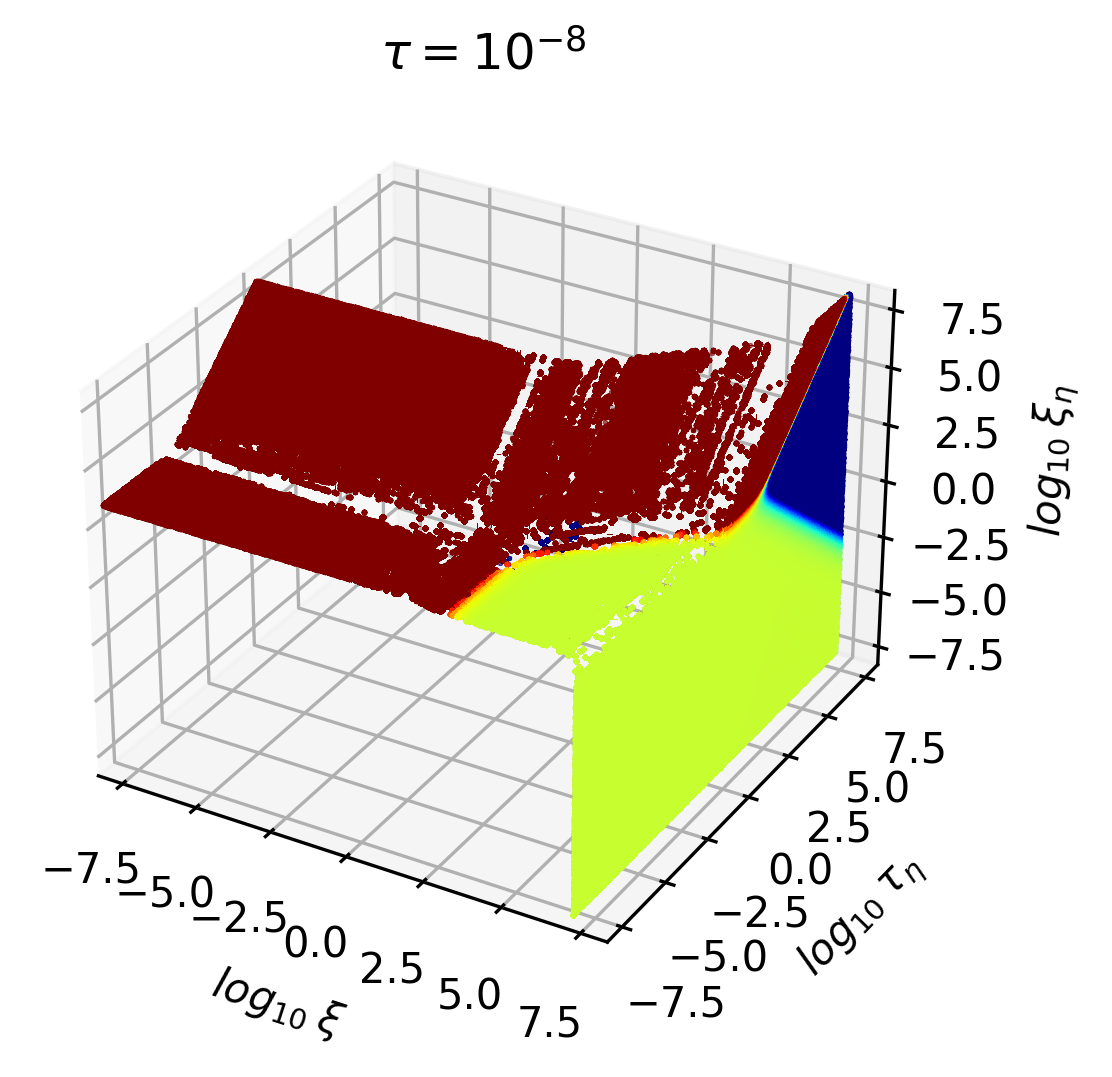} \\(a)}
             
        \end{minipage}}
               \fbox{\begin{minipage}[ht]{0.42\linewidth}
            \center{\includegraphics[width=\textwidth]{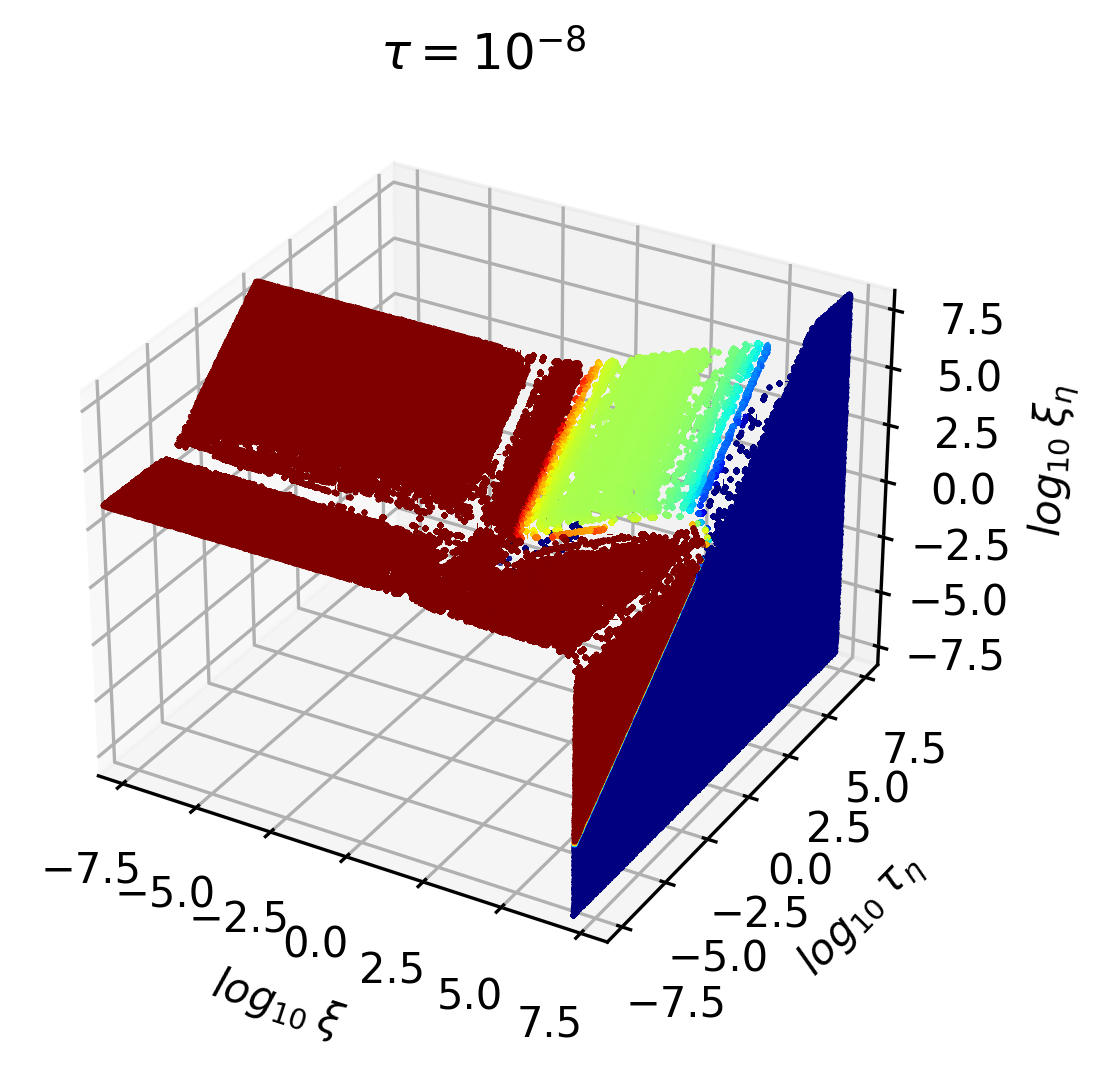} \\(b)}
             \end{minipage}}
                  \begin{minipage}[ht]{0.42\linewidth}
            \center{\includegraphics[width=\textwidth]{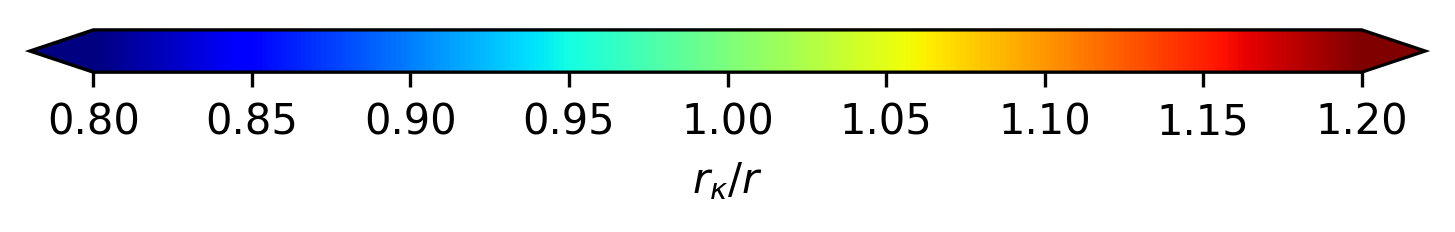}}
             \end{minipage}
                \begin{minipage}[ht]{0.42\linewidth}
            \center{\includegraphics[width=\textwidth]{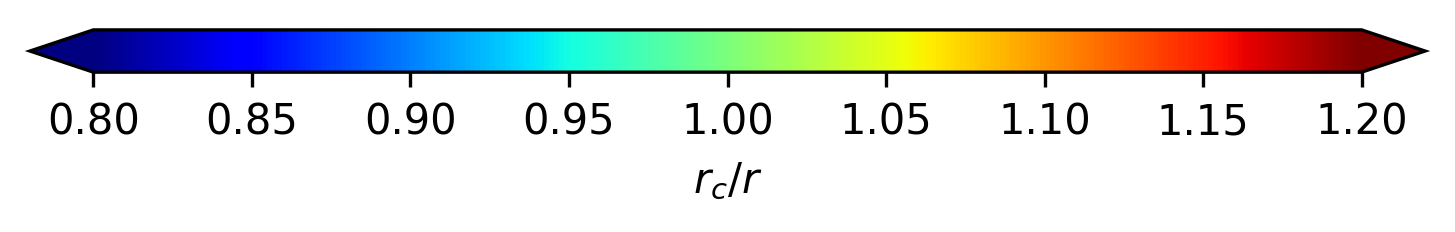}}
             \end{minipage}
    \caption{Comparison of the numerical value of the tensor-to-scalar ratio($r$) and the analytical prediction for the maximally symmetric case (left) and cubic pole (right) for a fixed value of $\tau=10^{-8}$. The quantity $r_\kappa$ is computed from Eq.~\eqref{ns-r-poles}, using the curvature in Eq.~\eqref{NY-curvature}.The quantity $r_c$ is computed from Eq~\eqref{eq:triple}. }
    \label{fig:NY-comparison}
    \end{center}
\end{figure}

%%%%%%%%%%%%%%%%%%%%%%%%%%%%%%%%%%%%%%%%%%%%%%%%%%%%%%%%%%%%%%%%%%%%%%%%%%%%%%%%
The results obtained by fixing $\tau=10^{-3}$ are shown in the panels (a) and (b) of Fig.~\ref{fig:NY-plots}. For the sake of completeness, we display also in Fig.~\ref{fig:runnings} the associated running and running of the running. We can distinguish several regimes. Firstly, for $\xi \ll 1$ the potential for the canonically normalized field becomes approximately quartic and the associated tensor-to-scalar ratio exceeds the observational bound. For $\tau_\eta \rightarrow 0$, we find a result analog to the one obtained in the single-field HI case~\cite{Langvik:2020nrs,Shaposhnikov:2020gts}, since in this limit the most relevant contribution to the inflationary parameters comes either from $\xi$ or $\xi_\eta$. In particular we can recognize a region of Palatini-like inflation for $\xi \sim 10^{7}$ and $\xi_\eta <\mathcal{O}(10^{3}) $, as well as a metric-like one for $\xi \sim\xi_\eta \sim 10^{3}$. The spectral tilt $n_s$ lies within the $\text{68\% C.L.}$ \textit{Planck} 2018 constraint  $n_s=0.9625 \pm 0.0048$ \cite{Planck:2018jri}, being well approximated by the corresponding expression in Eq.~\eqref{ns-r-poles}. 
 On the other hand, when $\tau_\eta$ increases, its contribution cannot be ignored any longer. For $\xi \sim 10^{7}$, we are once more in a Palatini-like regime, where the observables display only a mild dependence on $\tau_\eta$ and $\xi_\eta$. Conversely, for $1\ll \xi < 10^7$, the spectral tilt deviates clearly from \eqref{ns-r-poles}, rather approaching the cubic pole expectation in Eq~\eqref{eq:triple}. The panels (c) and (d) in Fig.~\ref{fig:NY-plots} illustrate how lowering the value of $\tau$ mildly affects the previous results. Indeed, we observe that the main features of panels (a) and (b) are overall maintained, being the main difference the small increase of the spectral tilt on the whole surface.
 %%%%%%%%%%%%%%%%%%%%%%%%%%%%%%%%%%%%%%%%%%%%%%%%%%%%%%%%%%%%%%%%%%%%%%%%%%%%%%%%
\begin{figure}
    \begin{center}
        \fbox{\begin{minipage}[ht]{0.42\linewidth}
            \center{\includegraphics[width=\textwidth]{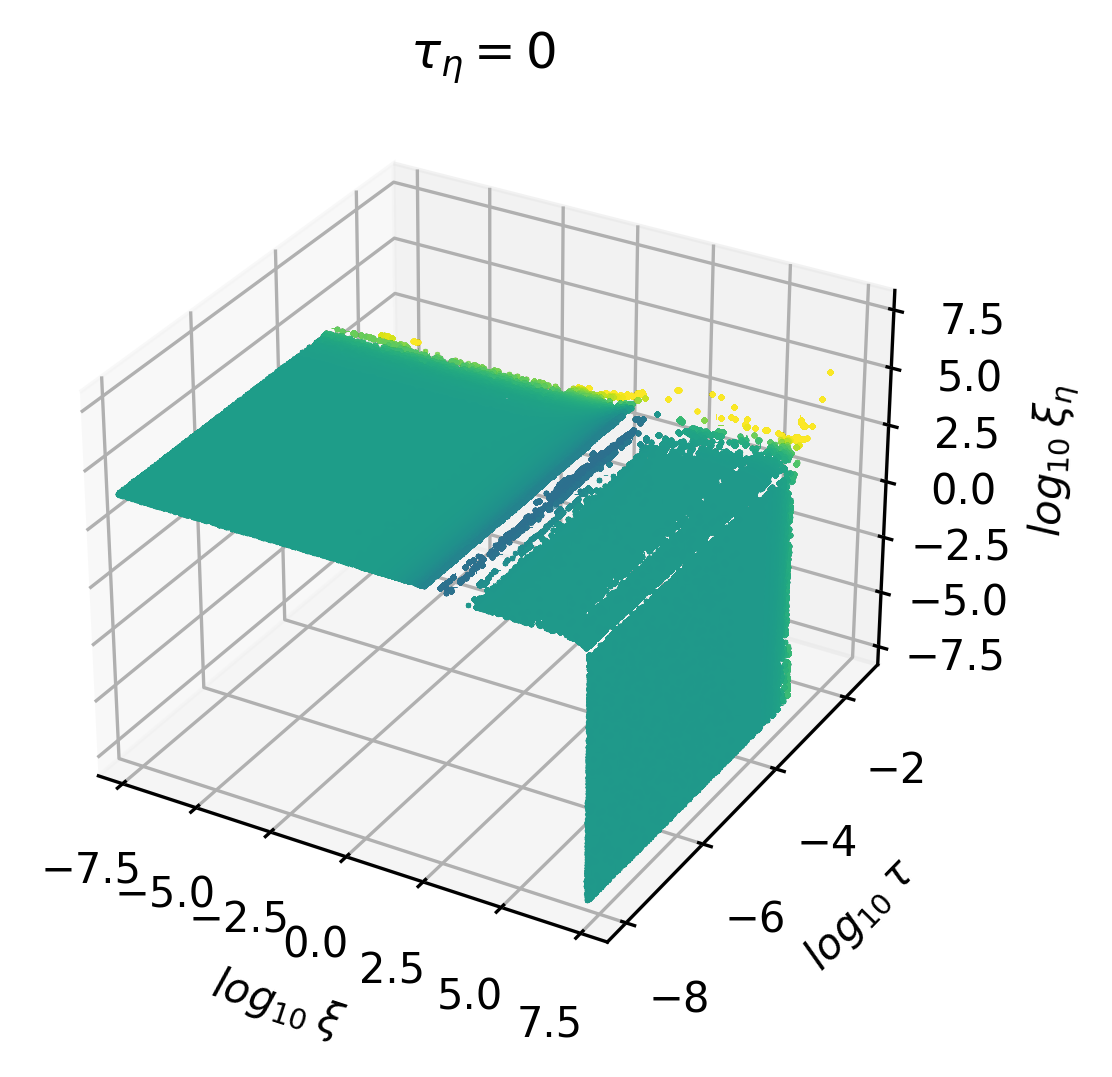} \\(a)}
        \end{minipage}}
        \fbox{\begin{minipage}[ht]{0.42\linewidth}
            \center{\includegraphics[width=\textwidth]{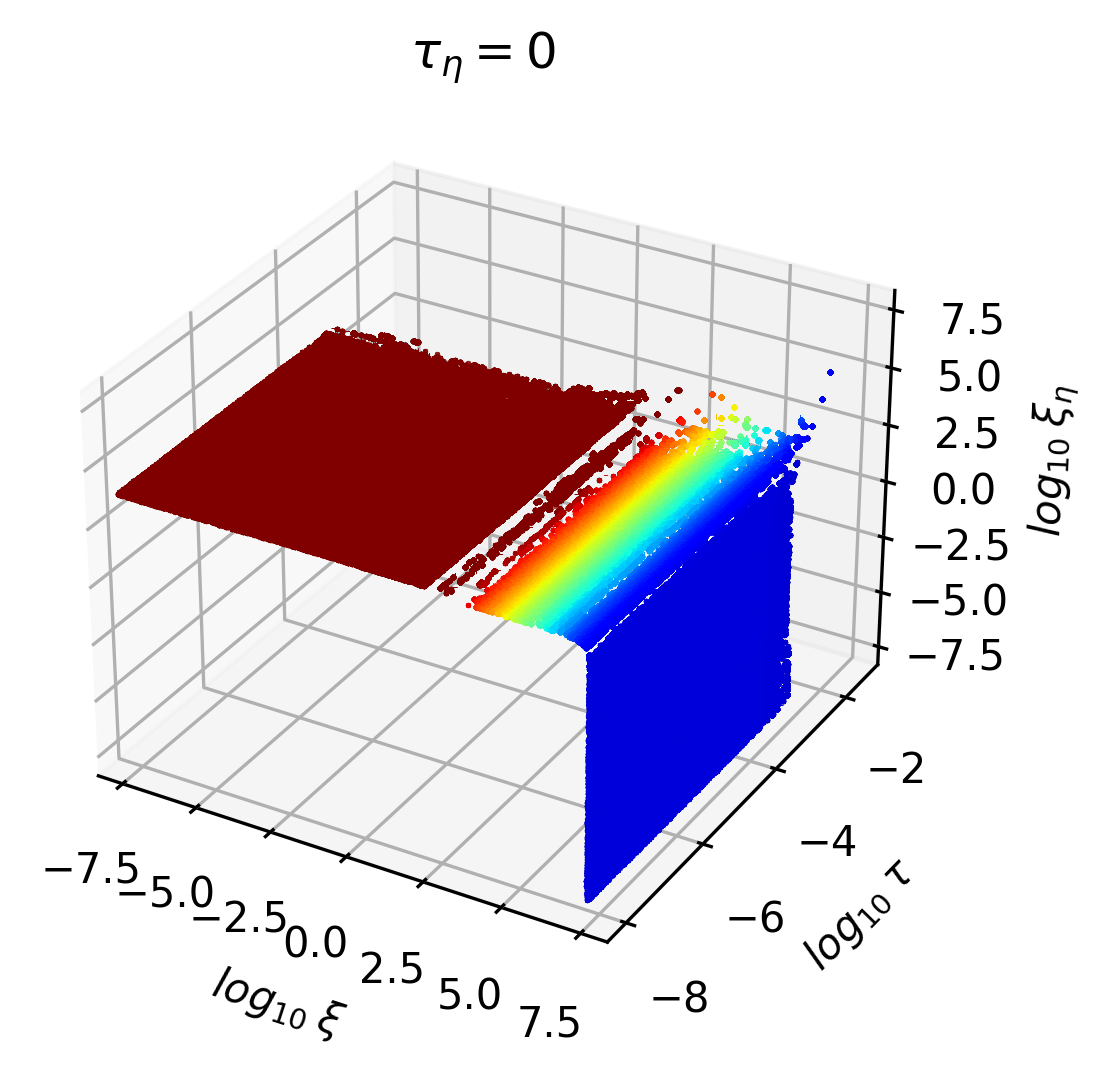} \\(b)}
        \end{minipage}}
        \begin{minipage}[ht]{0.42\linewidth}
            \center{\includegraphics[width=\textwidth]{Cbar_NS-std.png}}
             \end{minipage}
                \begin{minipage}[ht]{0.42\linewidth}
            \center{\includegraphics[width=\textwidth]{Cbar_R-std.png}}
             \end{minipage}
        \caption{Numerical results for the Nieh-Yan-like inflation case at vanishing $\tau_\eta$. The behaviour is substantially that of the single-field HI case. For $\tau > \mathcal{O}(10^{-2})$ the slow-roll conditions are quickly broken.}
    \label{fig:NY-SRB}
    \end{center}
\end{figure}
%%%%%%%%%%%%%%%%%%%%%%%%%%%%%%%%%%%%%%%%%%%%%%%%%%%%%%%%%%%%%%%%%%%%%%%%%%%%%%%%
 The dependence of $n_s$ on the value of $\tau$ is more clearly displayed in Fig.~\ref{fig:NY-SRB}, where we set $\tau_\eta=0$. Similarly to the Holst-like scenario, the spectral tilt decreases for larger values of $\tau$, saturating the observational bound at $\tau\sim 10^{-2}$.
In Fig.~\ref{fig:NY-comparison}  we compare the numerical results with the analytical predictions for the tensor-to-scalar ratio. Similarly to what happens for the spectral tilt, we observe that the maximally symmetric expectation in Eq.~\eqref{ns-r-poles}, with the formal replacement $\kappa_H\to \kappa_{NY}$, works well in the limit of small $\tau_\eta$. Conversely, for higher values of $\tau_\eta$ and $1\ll \xi < 10^7$, the behaviour of $r$ follows Eq.~\eqref{eq:triple}, being mainly driven by the cubic pole. Both approximations fail in the Palatini-like region where $\tau_\eta$ and $\xi_\eta$ are simultaneously large. This range of parameters corresponds to the transition between the two attractors in Fig.~\ref{fig:ns-r}, where the interplay between quadratic and cubic contributions cannot be longer ignored.

%%%%%%%%%%%%%%%%%%%%%%%%%%%%%%%%%%%%%%%%%%%%%%%%%%%%%%%%%%%%%%%%%%%%%%%%%%%%%%%%
%%%%%%%%%%%%%%%%%%%%%%%%%%%%%%%%%%%%%%%%%%%%%%%%%%%%%%%%%%%%%%%%%%%%%%%%%%%%%%%%

Lastly, we briefly discuss the case in which the two non-minimal couplings in the Nieh-Yan term are equal. As can be seen directly by imposing $\xi_\eta=\tau_\eta$ in Eq.~(\ref{NY-Ks}), we obtain  $a=0$ and therefore the same kinetic function as in Palatini HD inflation~\cite{Rubio:2020zht}, implying that this condition leads to the same inflationary phenomenology. A naive explanation for this behaviour can be inferred by looking at the Nieh-Yan action in the original frame (\ref{Nieh-Yan-Action}). Having the two non-minimal couplings equal leads to a term proportional to $\chi^2+h^2$. In Palatini HD inflation the physical dilaton $\rho$ is indeed a function of this quantity, as in Eq.~\eqref{Eq:Holst-diagonal}. This means that in this particular case the Nieh-Yan term modifies only the dynamics of the dilaton field, not affecting though that of the inflaton counterpart, which is regulated only by the $S_0$ action, thus originating pure Palatini inflation. Note, however, that with the condition $\tau_\eta=\xi_\eta$ the curvature in Eq.~(\ref{NY-curvature}) does not reduce to \eqref{eq:kH}. This implies that despite inflation takes place in a non-maximally symmetric region of the field manifold, we can still obtain observables in agreement with observations.

%%%%%%%%%%%%%%%%%%%%%%%%%%%%%%%%%%%%%%%%%%%%%%%%%%%%%%%%%%%%%%%%%%%%%%%%%%%%%%%%
\section{Discussion and outlook}\label{sec:conclusions}
%%%%%%%%%%%%%%%%%%%%%%%%%%%%%%%%%%%%%%%%%%%%%%%%%%%%%%%%%%%%%%%%%%%%%%%%%%%%%%%%

In this paper we have discussed the main features of Higgs-Dilaton inflation in the framework of Einstein-Cartan gravity, focusing our attention on the impact of non-minimal couplings to the Holst and Nieh-Yan terms. The resulting action contains four extra parameters on top of the two already appearing in the prototypical metric and Palatini formulations of the model. 

After transforming the theory to an equivalent metric formulation trading torsion for higher-dimensional operators within the Einstein-frame kinetic sector, we showed that the field-space metric displays a maximally symmetric structure at large values of the Higgs field and small dilaton couplings to the scalar curvature. Taking into account the conservation current of dilatations, we further reduced this manifold to a diagonal form, making explicit the physical dilaton field and reducing the apparently multifield inflationary dynamics to that of a single field scenario. Using then a MCMC method we determined the viable parameter space in the Holst- and Nieh-Yan-like limits and compared it to the analytical expectations following from the pole structure of the inflaton kinetic term.

The inflationary observables in Holst-like inflation were shown to reduce to those of pure Palatini HD inflation when the field-space curvature is approximately constant. Besides this behaviour, we also found that inflation can potentially take place in non-maximally symmetric regimes, albeit in those cases the tensor-to-scalar ratio exceeds the most recent observational bounds. The analytical and numerical treatment of Nieh-Yan-like inflation revealed the existence of two attractor solutions associated respectively to quadratic and cubic poles in the inflaton kinetic term and compatible with observations. The quadratic limit is again related to the emergence of a maximally symmetric regime and works well for small $\tau_\eta$ and large field-space curvature. On the other hand, the cubic pole becomes dominant at large values of $\tau_\eta$. This constitutes a unique feature of the Higgs-Dilaton scenario as compared with its single-field counterpart. It would be interesting to explore the implications of this large dilaton coupling for the validity of the effective field theory and the heating stage following the end of inflation.

All in all, it is fascinating that the inflationary predictions boil down to the geometrical properties of the field manifold for a sizable fraction of the parameter space. Interestingly enough, the associated scalar curvature is expected to play also a leading role in the attempt to generate a large hierarchy between  the electroweak and the Planck scale via gravitational instanton configurations \cite{Shaposhnikov:2018xkv,Shaposhnikov:2018jag,Shkerin:2019mmu,Karananas:2020qkp}. Indeed, as pointed out in Ref.~\cite{Karananas:2020qkp}, this quantity controls generically the vacuum expectation value of the Higgs field in the conformal SM limit $\alpha=0$,
\begin{equation}\label{eq:vev} 
\langle h \rangle \sim M_{P} e^{-{\cal W}(|\kappa|)} \ , 
\end{equation}
with ${\cal W}(|\kappa|)\sim {\cal O}(\sqrt{\vert \kappa\vert})$ a finite function. Large values of $\vert\kappa\vert$ may lead therefore to a small ratio $\langle h\rangle/M_P$, making \textit{a priori} possible to infer the value of the Fermi scale from CMB observations.

The analysis presented in this work constitutes a preliminary step towards a more accurate study taking into account the following issues:
\begin{enumerate}
\item \textit{Negative couplings}: We have restricted ourselves to positive non-minimal couplings, being this the simplest possibility to ensure the absence of negative eigenvalues in the field-space metric~\eqref{K-matrix}. Relaxing this assumption might lead to the appearance of new additional inflationary solutions, as those found for instance in Ref.~\cite{Langvik:2020nrs}, where the parameters of the single-field HI scenario were left unrestricted. 
\item \textit{Matter content}: We fixed the duration of inflation to  $N=55$ $e$-folds, ignoring the potential changes of the heating stage as a function of the model parameters. The associated uncertainties could be easily incorporated by increasing the value of $\sigma$ in Eq.~\eqref{likelihood}, resulting effectively in a mere enlargement of the available parameter space and a small scattering of points around the attractor solutions displayed in Fig.~\ref{fig:ns-r}. 

As pointed out in the introduction, the fermions in the SM and beyond can also modify the dynamics of the theory through non-minimal kinetic terms, manifesting themselves as higher-dimensional Higgs-fermion and fermion-fermion interactions in the equivalent metric formulation~\cite{Freidel:2005sn,Alexandrov:2008iy,Diakonov:2011fs,Shaposhnikov:2020aen}. While assumed to be negligible in this work, these terms come together with arbitrary coupling constants that could alter the inflationary energy distribution~\cite{Shaposhnikov:2020gts}, spoil the flatness of the potential, or induce new depletion channels beyond those already considered in the standard treatments of preheating~\cite{Garcia-Bellido:2008ycs,Bezrukov:2008ut,Bezrukov:2014ipa,Rubio:2015zia,Repond:2016sol,Ema:2016dny,DeCross:2016cbs,Sfakianakis:2018lzf,Rubio:2019ypq}. Besides, the four-legs fermion-fermion interactions could mediate the production of feebly interacting species which, if singlets under the SM gauge group, could play the role of dark matter \cite{Asaka:2005an,Asaka:2005pn}. This scenario has already been studied in Ref.~\cite{Shaposhnikov:2020aen}, with a focus on the observational consequences for right-handed neutrinos and their primordial momentum distribution. In the context of the HD model, the presence of extra couplings could lead to differences in the predictions, potentially modifying the viable parameter space.

\item \textit{Quantum corrections}: Our analysis is purely classical, not accounting therefore for the running of the Higgs quartic coupling or plausible threshold effects associated to the non-renormalizability of the Standard Model non-minimally coupled to gravity \cite{Bezrukov:2014ipa,Shaposhnikov:2020fdv}. The renormalization group running is expected to affect the shape of the inflationary potential in some corners of the parameter space, leading potentially to critical scenarios involving inflection points or wiggles along the inflationary trajectory and a broad bump in the power spectrum of curvature fluctuations at small and intermediate scales \cite{Rubio:2014wta}. Although this non-monotonic behaviour has been advocated to facilitate the formation of primordial black holes in the metric formulation \cite{Ezquiaga:2017fvi}, it unavoidably requires an unrealistic running of the Higgs non-minimal coupling to gravity significantly exceeding the 1-loop renormalization group result \cite{Rubio:2014wta}. It would be interesting to revisit this conclusion in the context of EC HI inflation, where the renormalization group equations are still unknown. The same applies to the impact of threshold effects, which depends implicitly on the effective cut-off scale of the theory, a quantity that can be significantly  different for distinct areas of the parameter space. 
\end{enumerate}
Further extensions within the general framework presented in Ref.~\cite{Karananas:2021gco} are also worthy to explore, at the expense of potentially reducing the model predictivity due to the many operators and functions involved. On top of that, our setting could be extended to include specific combinations of curvature-squared invariants resulting in healthy particle spectra, along the lines of Refs.~\cite{Gorbunov:2012ns,Ema:2017rqn,Gundhi:2018wyz,Gorbunov:2018llf,Edery:2018jyp,Ema:2019fdd,Ferreira:2019zzx,Ferreira:2021ctx,Ema:2020zvg,Blagojevic:2018dpz,Lin:2018awc,Karananas:2021zkl}.

%%%%%%%%%%%%%%%%%%%%%%%%%%%%%%%%%%%%%%%%%%%%%%%%%%%%%%%%%%%%%%%%%%%%%%%%%%%%%%%%
\acknowledgments
%%%%%%%%%%%%%%%%%%%%%%%%%%%%%%%%%%%%%%%%%%%%%%%%%%%%%%%%%%%%%%%%%%%%%%%%%%%%%%%%

We thank Georgios K. Karananas for useful discussions and comments on the manuscript. We acknowledge the Funda\c c\~ao para a Ci\^encia e a Tecnologia (FCT), Portugal, for the financial support to the Center for Astrophysics and Gravitation-CENTRA, Instituto Superior T\'ecnico,  Universidade de Lisboa, through the Project No.~UIDB/00099/2020.  M.~P. thanks  also the support of this agency through the Grant No. SFRH/BD/151003/2021 in the framework of the Doctoral Program IDPASC-Portugal. J.~R. thanks the Fundação para a Ciência e a Tecnologia (Portugal) for financial support through the CEECIND/01091/2018 grant. 

%%%%%%%%%%%%%%%%%%%%%%%%%%%%%%%%%%%%%%%%%%%%%%%%%%%%%%%%%%%%%%%%%%%%%%%%%%%%%%%%
\appendix
%%%%%%%%%%%%%%%%%%%%%%%%%%%%%%%%%%%%%%%%%%%%%%%%%%%%%%%%%%%%%%%%%%%%%%%%%%%%%%%%

%%%%%%%%%%%%%%%%%%%%%%%%%%%%%%%%%%%%%%%%%%%%%%%%%%%%%%%%%%%%%%%%%%%%%%%%%%%%%%%%
\section{Derivation of the Holst and Nieh-Yan terms from the full theory}\label{appendix:Derivation}
%%%%%%%%%%%%%%%%%%%%%%%%%%%%%%%%%%%%%%%%%%%%%%%%%%%%%%%%%%%%%%%%%%%%%%%%%%%%%%%%

Depending on the requirements chosen to build the action, the EC HD theory can accommodate a plethora of non-minimal operators with \textit{a priori} arbitrary coupling constants. A phenomenologically viable option was proposed in Ref.~\cite{Karananas:2021gco}, where the authors required the absence of extra gravitational degrees of freedom beyond the massless graviton and accounted only for operators with dimension not greater than four. The resulting theory entails still numerous operators and couplings, 
\begin{equation}
\begin{aligned}
\label{eq:action_general_unitary_scale}
S_{\rm SI} &= \int d^4 x \sqrt{g}\Bigg[\frac{\Omega^2}{2}{R} - \frac {(\partial_\mu \chi)^2}{2} -  \frac {(\partial_\mu h)^2}{2} -U(h,\chi) +  J^v_\mu v^\mu+  J^a_\mu a^\mu\\
&+ \frac{\tau\chi^2}{2}\left( G_{vv} v_\mu v^\mu  + 2G_{va}v_\mu a^\mu + G_{aa}	a_\mu a^\mu+G_{\tau\tau}\tau_{\alpha\beta\gamma} \tau^{\alpha\beta\gamma} + \tilde{G}_{\tau\tau} \epsilon^{\mu \nu \rho \sigma} \tau_{\lambda\mu\nu} \tau^\lambda_{~\rho\sigma}\right)   \Bigg]\,.
\end{aligned}
\end{equation}
In this expression, the kinetic and potential terms for the $h$ and $\chi$ fields, together with the coupling to the scalar curvature $R$, are in a one-to-one correspondence with those in Eq.~\eqref{EH-action}. On the other hand, the quantities 
\begin{equation}
\label{eq:tors_all}
v_\mu = T^\nu_{~\mu\nu}\,, \hspace{10mm}
a_\mu = \epsilon_{\mu\nu\rho\sigma}T^{\nu\rho\sigma}\,,\hspace{10mm}
\tau_{\mu\nu\rho} =\frac 2 3 \left( T_{\mu\nu\rho} -v_{[\nu} g_{\rho]\mu} - T_{[\nu\rho]\mu} \right)\,,
\end{equation}
stand for the irreducible components of the torsion tensor $T^{\nu\rho\sigma}$, 
\begin{equation}
    \label{Eq:GIJ}
    G_{AB}(\chi,h)=G_{AB}\left(\frac{h}{\chi}\right)=c_{AB}\left(1+\frac{\xi_{AB} h^2}{\tau \chi^2} \right)\,,
\end{equation}
are functions of the ratio $h/\chi$ with $A={v,a}$ and
\begin{equation}
    \label{Eq:Jva}
    J^{A}_\mu(\chi,h)=\tau_{A} \partial_\mu \chi^2+ \xi_{A} \partial_\mu h^2.
\end{equation}
Notice that in Eq~\eqref{Eq:GIJ} we do not consider the functions $G_{\tau\tau}$ and $\tilde{G}_{\tau\tau}$. The reason is that the tensorial part of torsion vanishes on the equations of motion, and therefore all results are independent of these quantities~\cite{Karananas:2021gco}.
After solving for torsion and moving to the Einstein frame, the total action \eqref{eq:action_general_unitary_scale} can be written as
\begin{equation}
\begin{aligned}
\label{eq:action_gen_Einst_scale}
S_{\rm SI}&=\int d^4 x \sqrt{g} \Bigg[\frac{R}{2} -\frac {1} {2\Omega^2} \widetilde \gamma_{ij} g^{\mu\nu}\partial_\mu \varphi_i \partial_\nu \varphi_j  - \frac{U(h,\chi)}{\Omega^4}\Bigg] \ ,
\end{aligned}
\end{equation}
with 
\begin{equation}
\widetilde \gamma_{ij} = \mathcal I_{ij}+\frac{4}{\tau\chi^2\left(G_{vv}G_{aa}-G_{va}^2\right)}\gamma_{ij}+\frac {6}{\tau \chi^2+\xi h^2}
\begin{pmatrix}
\tau^2\chi^2&\tau\xi\chi h \\
\tau\xi\chi h&\xi^2h^2
\end{pmatrix}\,,
\end{equation}
the metric of the two-dimensional field manifold spanned by $\varphi^i=(\chi,h)$, $\mathcal I_{ij}$ the $2\times 2$ identity matrix and
\begin{eqnarray}
\gamma_{11}&=&\left(G_{aa}\tau_v^2+G_{vv}\tau_a^2-2G_{va}\tau_v \tau_a\right)\chi^2 \, ,\\
\gamma_{12}&=&\left(G_{aa}\xi_v\tau_v+G_{vv}\xi_a\tau_a-G_{va}\left(\xi_v \tau_a+\tau_v\xi_a\right)\right)h\chi \,,  \\
\gamma_{22}&=&\left(G_{aa}\xi_v^2+G_{vv}\xi_a^2-2G_{va}\xi_v \xi_a\right)h^2  \, .
\end{eqnarray}
The action \eqref{Eq:Final-Action} for the HD model with the Holst and Nieh-Yan term follows from the choice
\begin{eqnarray}
 c_{vv}&=&-\frac{2}{3}\,,\hspace{6mm} c_{va}=\frac{1}{6 \bar{\gamma}}\,,\hspace{6mm} c_{aa}=\frac{1}{24}\,,\hspace{6mm}\xi_{vv}=\xi_{aa}=-\xi_{v}=\xi\,,\hspace{6mm}\xi_{va}=\xi_\gamma\,,\\
 \xi_{a}&=&\frac{\xi_\gamma}{4 \bar{\gamma}}+\frac{\xi_\eta}{4}\,,\hspace{6mm}\tau_{a}=\frac{\tau}{4\bar{\gamma}}+\frac{\tau_\eta}{4}\,,\hspace{6mm}\tau_{v}=-\tau\,.
\end{eqnarray}

%%%%%%%%%%%%%%%%%%%%%%%%%%%%%%%%%%%%%%%%%%%%%%%%%%%%%%%%%%%%%%%%%%%%%%%%%%%%%%%%
\section{Diagonalization of Nieh-Yan Lagrangian}\label{appendix:diagonalization}
%%%%%%%%%%%%%%%%%%%%%%%%%%%%%%%%%%%%%%%%%%%%%%%%%%%%%%%%%%%%%%%%%%%%%%%%%%%%%%%%

In this appendix we show explicitly how to diagonalize the Nieh-Yan-like field-metric (\ref{NY-kin}) using the conservation of the Noether current associated to scale invariance. In a covariant formulation, this reads
\begin{equation}\label{cov-current}
    J_\mu = -K_{ab}\partial_\mu \varphi^a \Delta \varphi^b\,, 
\end{equation}
with $\Delta \varphi^a$ the variations of the fields under infinitesimal dilatations.~\footnote{Note that, taking into account the explicit form of this expression in $h$ and $\chi$ variables ($\Delta h=h$, $\Delta\chi=\chi$),
\begin{equation}\label{current}
    J_\mu= \frac{-1}{2\Omega^2} \left[ \partial_\mu (\chi^2 + h^2) + 6 \frac{\tau_\eta \chi^2 + \xi_\eta h^2}{\Omega^2}\partial_\mu (\tau_\eta\chi^2 + \xi_\eta h^2)\right]\,, \nonumber
\end{equation}
we can easily recover the metric and Palatini cases in the limits $\tau_\eta=\tau,\;\;\xi_\eta=\xi$ and $\tau_\eta=\xi_\eta=0$ respectively~\cite{Rubio:2020zht}.}
 Following the procedure outlined in Ref.~\cite{Casas:2018fum}, we look for a change of variables $(\chi,h)\rightarrow (\rho,\theta)$ able to bring the current to the form 
\begin{equation}
    J_\mu=M_P K_{\rho \rho}(\theta) \partial_\mu\, \rho\,.
\end{equation}
To this end, we start by choosing a generic change of variables
\begin{equation}\label{variables}
    \tilde{\rho}= \frac{1}{2} \ln \left(\chi^2 + h^2\right)\,, 
    \hspace{15mm}
    \theta= \arctan\left( \frac{h}{\chi}\right)\,,
\end{equation}
leading to a structure
\begin{equation}
    J_\mu=\frac{-1}{2f_\Omega}\left[2\left(1+6\frac{f_\eta^2}{f_\Omega} \right)\partial_\mu \tilde{\rho} + 6 \frac{f_\eta}{f_\Omega}\frac{\partial f_\eta}{\partial\theta}\partial_\mu \theta\right]\equiv  -K_{\tilde{\rho}\tilde{\rho}}(\theta) \partial_\mu \tilde{\rho} -K_{\theta \theta}(\theta)\partial_\mu \theta\,,
\end{equation}
with
\begin{equation}
    f_\Omega (\theta)= \tau \cos^2\theta + \xi \sin^2\theta\,, \hspace{15mm}
    f_\eta (\theta)= \tau_\eta \cos^2\theta + \xi_\eta \sin^2\theta\,,
\end{equation}
two new functions whose ratio parametrize the departure from the metric limit. In order to get rid of the undesired $\partial_\mu \theta$ term, we define now a shifted radial variable 
\begin{equation}\label{new-rad0}
    \rho=\tilde{\rho} + \int^\theta d\theta' \frac{K_{\theta \theta}(\theta')}{K_{\tilde{\rho}\tilde{\rho}}(\theta')}\,,
\end{equation}
reading explicitly 
\begin{equation}\label{new-rad}
    \rho=\tilde{\rho}+\frac{1}{4} \int^\theta d\theta' \frac{6}{f_\Omega + 6 f^2_\eta}\frac{\partial f^2_\eta}{\partial \theta'} =\tilde{\rho}+\frac{1}{4} \ln (f_\Omega + 6f_\eta^2) - \frac{1}{4} \int^\theta d\theta' \frac{1}{f_\Omega + 6f_\eta^2}\frac{\partial f_\Omega}{\partial \theta'}\,.
\end{equation}
Taking into account that
\begin{equation}
    \begin{split}
      &\int^\theta d\theta' \frac{1}{f_\Omega + 6f_\eta^2}\frac{\partial f_\Omega}{\partial \theta'} =
       \frac{\xi -\tau }{A}\ln \Big\vert  \frac{A-B}{A+B}\Big\vert + \ln C
       \,,
    \end{split}
\end{equation}
with $C$ an arbitrary integration constant and  
\begin{equation}
    \begin{split}
    B&=\xi-\tau+12 \left[( \xi _{\eta } \tau _{\eta }- \tau _{\eta }^2 )\cos^2 \theta+
(\xi _{\eta }^2-\xi _{\eta } \tau _{\eta }) \sin ^2\theta  \right]\,,\\ 
A^2&=(\xi -\tau )^2+24 \left[ ( \xi _{\eta } \tau _{\eta }  -  \tau _{\eta }^2)\xi - ( \xi _{\eta }^2-\xi _{\eta } \tau _{\eta })\tau \right]\,,
    \end{split}
\end{equation}
we obtain the following expression for the physical dilaton $\rho$ 
\begin{equation}\label{rad-var}
    \rho=\tilde{\rho} + \frac{1}{4}\ln  \left[C \Big\vert  \frac{A+B}{A-B}\Big\vert^{\frac{\xi -\tau }{A}} (f_\Omega + 6f_\eta^2)\right] \,.
\end{equation}
Replacing the resulting pair $(\theta,\rho)$ into the Nieh-Yan action, we recover the kinetic functions in Eqs.~\eqref{NY-Ks} and \eqref{eq:Krho}. 

%%%%%%%%%%%%%%%%%%%%%%%%%%%%%%%%%%%%%%%%%%%%%%%%%%%%%%%%%%%%%%%%%%%%%%%%%%%%%%%%
\bibliographystyle{JHEP}
\bibliography{mybib}
%%%%%%%%%%%%%%%%%%%%%%%%%%%%%%%%%%%%%%%%%%%%%%%%%%%%%%%%%%%%%%%%%%%%%%%%%%%%%%%%
\end{document}